\documentclass[lineno]{biometrika}

\usepackage{amsmath}
\usepackage{url}

\usepackage{newtxtext}
\usepackage[subscriptcorrection]{newtxmath}
\usepackage{subfiles}

\usepackage [autostyle, english = american]{csquotes}
\MakeOuterQuote{"} 

\graphicspath{{./art/}}

\usepackage[plain,noend]{algorithm2e}
\usepackage{float}

\makeatletter
\renewcommand{\algocf@captiontext}[2]{#1\algocf@typo. \AlCapFnt{}#2} 
\def\@algocf@capt@plain{top}
\renewcommand{\algocf@makecaption}[2]{%
  \addtolength{\hsize}{\algomargin}%
  \sbox\@tempboxa{\algocf@captiontext{#1}{#2}}%
  \ifdim\wd\@tempboxa >\hsize
    \hskip .5\algomargin%
    \parbox[t]{\hsize}{\algocf@captiontext{#1}{#2}}
  \else%
    \global\@minipagefalse%
    \hbox to\hsize{\box\@tempboxa}
  \fi%
  \addtolength{\hsize}{-\algomargin}%
}
\makeatother

\usepackage{dsfont}
\usepackage{bm}
\usepackage{enumitem}
\newcommand\ci{\perp\!\!\!\perp}
\newcommand\obs{\mathcal{O}}
\newcommand\prp{pr}
\newcommand\e{E}

\usepackage{tikz}
\usepackage{accents}
\usetikzlibrary{calc}
\usetikzlibrary{graphs}
\usetikzlibrary{positioning}
\usetikzlibrary{arrows}
\usetikzlibrary{automata}
\usetikzlibrary{shapes.geometric}
\usetikzlibrary{shapes}
\usetikzlibrary{arrows.meta}
\usetikzlibrary{chains,matrix,scopes}
\usetikzlibrary{backgrounds}

\newcommand*\diff{\mathop{}\!\mathrm{d}}

\nolinenumbers

\begin{document}

\jname{arXiv.org e-Print archive}
\jyear{2026}
\jvol{Volume}
\jnum{Number}
\cyear{2026}


\markboth{Maltzahn et~al.}{Robust estimation of occupation probabilities for coarsened multistate processes}

\title{Robust estimation of occupation probabilities for coarsened multistate processes}

\author{Niklas Nyboe Maltzahn}
\affil{Department of Biostatistics, University of Oslo,\\ 
P.O.Box 1122 Blindern, 0317 Oslo, Norway
\email{n.n.maltzahn@medisin.uio.no}}

\author{Gergely Dániel Lukáts}
\affil{Department of Biostatistics, University of Oslo,\\ P.O.Box 1122 Blindern, 0317 Oslo, Norway
\email{g.d.lukats@medisin.uio.no}}

\author{Kjetil Røysland}
\affil{Department of Biostatistics, University of Oslo,\\ P.O.Box 1122 Blindern, 0317 Oslo, Norway
\email{kjetil.roysland@medisin.uio.no}}

\maketitle


\begin{abstract}
    We derive augmented inverse probability weighted estimators for occupation probabilities of multistate models under two levels of coarsening; right-censoring and baseline exposure. The key exchangeability assumption for identification is coarsening at random, while allowing for time-varying confounders, but not requiring Markov properties. Using existing techniques from causal inference and missing data literature, the derived estimators have highly desirable robustness and efficiency properties. These properties are demonstrated through both theoretical results, and a simulation study.
\end{abstract}

\section{Introduction}


From cancer remission to job turnover, many phenomena can be described by how an entity moves between well-defined states over time. Statistical models of multistate processes, multistate models, offer a general framework for representing such state transition dynamics and encompass well-known cases such as survival (or time-to-event) models, competing events models, and recurrent events models. The flexibility of these models has made them useful across many disciplines and they have been used to model, for example, marital or employment status of an individual, credit rating and default status of a company, or warranty status of a product for determining premiums. For additional examples from specific fields, see \citet{Hoem1976} for demography; \citet{NEYMAN51}, \citet{Keiding2001}, and \citet{Keiding2002} for medical statistics; and \citet{MC2012} for actuarial science.


Given a multi-state process, a typical parameter of interest is the \textit{occupation probability}, i.e., the probability that the process is in a particular state, as a function of time. A standard example is the survival function, the probability of staying in the initial "alive" state, in a two state model of time-to-death. Another important parameter is the \textit{expected length of stay} (ELOS) in a given state, which is the occupation probability integrated over a time interval \citep{Hein2016, Hoffe22}. 

In this paper, we show how methods from the causal inference and missing data literature can be used to identify and estimate occupation probabilities of counterfactual multistate processes coarsened by baseline exposure and right censoring. We identify such parameters under weaker assumptions than what is commonly seen in the literature on multistate processes, both on the counterfactual multistate processes and the coarsening mechanisms. More specifically, we do not assume that the counterfactual multistate processes are Markov processes, and we allow the right censoring mechanism to depend on both the natural history of the process itself and other time dependent processes, relying on the assumption that the coarsening mechanisms are coarsening at random (CAR). Under stated identifiability criteria, we identify the counterfactual occupation probabilities using an inverse probability weighted (IPW) estimand and derive its efficient influence function. Based on the representation of the efficient influence function, we suggest, inspired by earlier work of \citet{Hubbard2000}, five different estimators. In particular, the one-step estimator exhibits better robustness and efficiency properties than a simple IPW-estimator. We discuss the large sample properties of the one-step estimator and compare the performance of the suggested estimators under correct specification or misspecification of nuisance parameter estimators. While similar estimation methods have been discussed before, we emphasize that with our weak assumption criteria, the proposed estimators apply broadly to multistate models with an arbitrary number of states, and allow state transitions and time-varying covariates to be non-Markov processes, unlike previous methods which are tailored to specific models under various restrictive criteria. See e.g., \citep{Martinussen2023} for methods on the illness-death models without recurrence, \citep{Scheike2020, Rytgaard2023} for competing events models, and \citep{baer2023causal, Miloslavsky2004} for recurrent events models.

In Section \ref{sec:EffInfl}, we describe the data structure, define the parameter of interest (counterfactual occupation probabilities) and state our identifiability criteria. We also define the identifying IPW estimand and derive its efficient influence function. In Section \ref{sec:est_occ_prob}, we define five different estimators—including the one-step estimator, for which we discuss large sample properties. In Section \ref{sec:4:simulations}, we compare the performance of the estimators in a simulation experiment and finally, in Section \ref{sec:discussion}, we summarise the main points of the paper.

\section{Efficient IF under right censoring and treatment assignment}
\label{sec:EffInfl}

\subsection{Data description, target and identifiability criteria}
\label{subsec:data}

To define the full data within intervention arms, for $a \in \{0,1\}$, let $Z_{a}$ be the latent data process, i.e., a càdlàg multistate process on the positive real line that takes values in $[K] := \{1, \ldots, K\}$. We can interpret $Z_{a}$ as representing e.g., an individual's states through time since inclusion into a study, under a hypothetical intervention $a$. Let $E_{abs} \subset [K]$ be the set of absorbing states, including, e.g., the state of death, and define the time to absorption $\tau_{a}:= \inf\{t : Z_{a}(t) \in E_{abs}\}$. Let $L_{a}$ be a time-dependent process of covariates, such as age, measured blood pressure or glucose levels, and let $W$, containing $L(0)$, be baseline measurements prior to treatment assignment and therefore, not dependent on $a$. To have some structure on $L_{a}$, assume it is likewise càdlàg and of bounded variation. Following the notation of \citet{LR2003}, we define for each arm, the data process  $X_a(u) := \{\mathds{1}\{\tau_{a} \leq u\}, Z_{a}(u), L_{a}(u)\}$, the data up until time $u$, $\bar{X}_{a}(u) := \{X_a(s): s \leq u\}$, and the (stopped) process history up to absorption $\bar{X}_{a} := \bar{X}_{a}(\tau_{a})$. Then, the full counterfactual data process is $\bar{\bm{X}} := (\bar{X}_{0}, \bar{X}_{1})$. We define $\bar{Z}_{a}(u)$ and $\bar{L}_{a}(u)$ analogously to $\bar{X}_{a}(u)$. Under a given intervention $a$, the target parameter, at fixed times $t_{0}$ and states $j$ are the (full-data) state occupation probabilities
\begin{align}
    \psi_{\bar{\bm{X}}}^{t_0, j}(F) := \e_{F}[\mathds{1}\{Z_{a}(\tau_{a} \wedge t_{0}) = j\}]. \label{eq:target}
\end{align}
Here, the superscript $F$ refers to the law of the full data process $\bar{\bm{X}}$. As a functional of F, this defines the full data estimand $\psi_{\bar{\bm{X}}}(F)$. Going forward, the dependency on $t_0$ and $j$ is often suppressed in notation, unless it is necessary for distinction. \\

The coarsening mechanism, which coarsens the full data to observed data, consists of two levels: A latent right-censoring time $C_{a}$ under the hypothetical assignment $a$, representing e.g., drop-out under intervention $a$, and a dichotomous observed treatment variable $A$. The two-dimensional random variable $(C_{A}, A)$ will be the \textit{coarsening variable} for the full data process $\bar{\bm{X}}$. Individuals are followed until either right-censoring or absorption. For technical reasons, to ensure that the coarsening variable is always observed, we demand that any law of interest satisfy that $C_A := \infty$ on events where $\{C_{A} \geq \tau_{A}\}$. The observed data is $\obs_A = (C_{A}, A,\Delta_{A}, \bar{Z}_{A}(\tau_A  \wedge C_A), \bar{L}_{A}(\tau_A  \wedge C_A))$, where $\Delta_{A} := \mathds{1}\{\tau_{A} \leq C_{A}\}$. To identify \eqref{eq:target}, we assume consistency and restrict attention to probability measures $\prp$, which (in addition to the always observed property of censoring) obey the following identifiability criteria: 

\begin{enumerate}[itemsep=10pt]
    \item \textit{Consistency assumption:} \\
    We assume that $A$ is a binary random variable that represents treatment allocation. The treatment levels $a$ index the counterfactual outcome processes $\bar X_a$. Additionally, there exists a mapping $\Phi$, which establishes that the observed data is a consistent coarsening of the full counterfactual data $\bar{\bm{X}}$ given the coarsening variable $(C_{A}, A)$. For individuals with $A=a$:
    \begin{align*}
        \Phi((C_{A}, A), \bar{\bm{X}}) &:= (C_{a}, a,\Delta_{a}, \bar{X}_{a}(\tau_a  \wedge C_a))
    \end{align*}
    
     Going forward, we omit the $A$ subscript for observed variables for brevity, i.e., $C := C_{A}$, and denote $\bar{X}(C) := \bar{X}_A(\tau_A \wedge C_A)$.
    \item \textit{Positivity assumption:} 
    \begin{align}
        \prp(A = a \vert W) \prp(C_{a} > t \vert \bar{X}_{a}) > 0, \; \prp \text{ a.s for a} \in \{0,1\} \label{eq:ObsPos}
    \end{align}
    
    \item \textit{Exchangability (CAR) assumption:} \\
    We assume a set of exchangability assumptions, implying that the coarsening mechanism satisfies CAR. For $a \in \{0,1\}$:
    \begin{align}
    \bar{X}_{a}, C_{a} &\ci_{\prp} A \ \vert \ W, \label{eq:CAR_A} \\
        \prp(A=a, C_{a} \in \diff t \vert \bar{\bm{X}}) &= \prp(A=a, C_{a} \in \diff t \vert \bar{X}_{a}), \text{ and} \label{eq:CAR_C1} \\
        \prp(A=a, C_{a} \in \diff t \vert \bar{X}_{a}) &= \prp(A=a, C_{a} \in \diff t \vert \bar{X}_{a}(t)). \label{eq:CAR_C2}
    \end{align}

\end{enumerate}

To see that the above assumptions imply CAR, refer to Appendix \ref{app:A:CAR}. These three criteria can be seen as latent event process analogues to the common three identifiability criteria from discrete time causal inference.

\subsection{Statistical model when coarsening is always observed}
\label{subsec:statmodel}

A consequence of the CAR assumption, combined with coarsening always observed, is a particular factorization of the observed data law, which suggests a particular model parameterization. We explain this model choice in pseudo formal notation. For more detail see e.g. \cite{GLR1997}, \cite{LR2003} or \cite{Tsiatis2006}. Due to assumptions \eqref{eq:CAR_A} to \eqref{eq:CAR_C2}, the coarsening kernel under fixed $\prp$ factorizes into conditional kernels of $A$ and $C_a$. We denote $\pi$ and $G$, in the following way:  
\begin{align}
    \prp(A=a, C_{a} \in \diff t \vert \bar{\bm{X}}) &= \prp(A = a \vert W)\prp(C_{a} \in \diff t \vert \bar{X}_{a}(t)) = \pi(a \vert W)G(\diff t \vert \bar{X}_{a}(t), a) .  \label{eq:censCoars}
\end{align}
Here, the kernel $\pi$ is indexed by the outcome space of $W$, and the kernel $G$ is indexed by the path space of $\bar{X}_{a}$ and $a$. Both can be taken to be zero outside the support of the law of $W$ and $\bar{X}_{a}$, respectively. Intuitively, this notation is also taken to mean that under fixed $\prp$, the decision to treat depends only on baseline covariates, and similarly, the infinitesimal probability of censoring at time $t$ does not depend on data after time $t$. By assumption \eqref{eq:CAR_A}, and consistency, we see that $G$ also determines the conditional law of the observable censoring $C_A$;
\begin{align}
    \prp(C_A \in \diff t \vert \bar{X}_{a}, A=a) =  G( \diff t \vert \bar{X}_{a},a), \quad \text{and} \quad \prp(C_A \in \diff t \vert \bar{X}_A,A) = G( \diff t \vert \bar{X}_A,A). \label{eq:ObsCFcens}
\end{align}
 Under the identifiability criteria above, the observed data distribution $P$ of $\obs$ under $\prp$ factorizes into (local) variation independent\footnote{See e.g., \cite{LR2003}, \cite{Tsiatis2006}} parts determined by $G$, $\pi$, and $F$, the law of $\bar{\bm{X}}$. This suggests a parameterization by the 3-tuple $\theta = (F,\pi,G)$, letting each of the parameters vary, up to identifiability. We will informally denote this set by $\Theta$. We then consider a statistical model for $\obs$ by the parameterization $\{\prp_{\theta} : \theta \in \Theta\}$ of measures on the sample space satisfying our identifiability criteria, and corresponding distributions $\{ P_{\theta} : \theta \in \Theta\}$ of $\obs$, such that $\obs$ has distribution $P_{\theta}$ under $\prp_{\theta}$. The true law of $\obs$ is determined by the true parameters $F_{0}, G_{0}$, and $\pi_{0}$, which we denote by $P_0$, the distribution of  $\obs$ under $\prp_0$. To avoid additional notation, $G$ will also denote the censoring survival function $G(t \vert \bar{X}(C),A) := \prp_{F,G, \pi}(C > t \vert \bar{X}(t \wedge C),A)$, but it will be clear from the arguments of $G$ whether we think of it as a function or a kernel. 

\subsection{Identification and efficient influence function}
\label{subsec:Indentification}

We identify the target \eqref{eq:target} over the assumed model by an observed data estimand, based on inverse probability weighting. To define this estimand, we introduce the $t_{0}$-censoring indicator
\begin{align*}
    \Delta_{a}^{t_{0}} := \Delta_{a} \vee \mathds{1}\{C_{a} \geq t_{0}\} = \mathds{1}\{\tau_{a} \wedge t_{0} \leq C_{a}\},
\end{align*}
which indicates whether an individual is censored no later than $t_{0}$ during follow-up. Define also the observed stopping time $\tilde{\tau} := \tilde{\tau}_{A} := \tau_{A} \wedge C_{A}$. This is an observed variable, since it is $\tau_A$ when $\Delta = 1$ and $C$ when $\Delta = 0$. Using these variables, and $G$ as a survival function, the observed data estimand is the mapping
\begin{align}
    \psi_{\obs}(F,G,\pi) &:= \e_{F,G, \pi}\left[\frac{\mathds{1}\{A = a\}\Delta^{t_{0}} \mathds{1}\{Z( \tilde{\tau} \wedge t_{0} ) = j\}}{\pi(a|W) G(\tilde{\tau} \wedge t_{0} \vert \bar{X}(C),A)}\right]. \label{eq:obs_estimand}
\end{align}
By the law of iterated expectations, conditional on $\bar{X}_{a}$, we see that
\begin{align*}
\psi_{\obs}(F,G,\pi) = \psi_{\bar{\bm{X}}}(F)
\end{align*}
i.e., we have identification. To derive the efficient influence function of the observed data estimand \eqref{eq:obs_estimand}, we use a general procedure which has been applied to many similar cases with similar coarsening structures, notably in \cite{Hubbard2000}. Again, we provide only a semi-formal description. For more details, see \cite{LaanHubbard1999}, \cite{LR2003}, \cite{Vaart2004} and \cite{Tsiatis2006}. Given the model $\{P_{\theta} : \theta \in \Theta\}$, the tangent space $\mathbb{T}_{\theta}(P_{\theta})$ of the model at the point $P_{\theta}$ can be defined as the closed set of scores defined through one-dimensional parametric sub-models $\{P_{F_{\varepsilon},G_{\varepsilon},\pi_{\varepsilon}}: \varepsilon\}$ passing through $P_{\theta}$. Generally, the tangent space $\mathbb{T}_{\theta}(P_{\theta})$ is a subspace of the $P_{\theta}$-zero mean function space $L^0_{2}(P_{\theta})$, but assuming CAR and the factorization \eqref{eq:censCoars}, these spaces are equal, and $\mathbb{T}_{\theta}(P_{\theta})$ is considered maximal. Furthermore, the tangent space $\mathbb{T}_{\theta}(P_{\theta})$ factorises as a direct sum of orthogonal parts
\begin{align*}
    \mathbb{T}_{\theta}(P_{\theta}) = \mathbb{T}_{F}(P_{\theta}) \oplus \mathbb{T}_{G}(P_{\theta}) \oplus \mathbb{T}_{\pi}(P_{\theta}),
\end{align*}
where each of the spaces in the direct sum are defined from scores of one dimensional sub-models through only $F$, $G$ or $\pi$, respectively. Here, the notion of orthogonality is defined with respect to the inner product in the $L^0_{2}(P_{\theta})$ space. In general, the part of the tangent space defined by the coarsening kernel, i.e., the left hand side of \eqref{eq:CAR_C1}, is denoted $\mathbb{T}_{CAR}$, and under our identifiability criteria, it further factorises as $\mathbb{T}_{CAR} = \mathbb{T}_{G}(P_{\theta}) \oplus \mathbb{T}_{\pi}(P_{\theta})$, following from \eqref{eq:censCoars}. Utilizing a similar decomposition of the score space, \citet{GLR1999} suggests a general procedure to derive the non-parametric efficient influence function for observed data estimands, such as our estimand in \eqref{eq:obs_estimand}. It consists of three steps: 
\begin{enumerate}
    \item Deriving the efficient score of a full data mapping, typically an expectation, such as \eqref{eq:target},
    \item Mapping it to the score space of $\mathbb{T}_{\theta}(P_{\theta})$ through inverse probability weighting, as in \eqref{eq:IF_0},
    \item Then, subtracting the orthogonal projection of the mapping to $\mathbb{T}_{CAR}$.
\end{enumerate}
The difficulty lies in identifying the projection in step three. Since our data structure is an expansion, and thus similar to that of \citet{Hubbard2000}, their projection formula can be leveraged, and a formal derivation of the projection is given Appendix \ref{app:B:inf_fn}. To present the influence function of $\psi_{\obs}$, we first introduce some notation for different components of the projection term. To that end fix $\theta = (F,G,\pi)$, and define from the kernel $G$, the hazard $\lambda^{G}$:
\begin{align}
    \lambda^{G}(t \vert \bar{X}, A)\diff t := \frac{G(\diff t \vert \bar{X}, A)}{G([t, \infty] \vert \bar{X}, A)} \label{eq:Ghaz_def}.
\end{align}
Under assumption \eqref{eq:censCoars}, $\lambda^{G}$ is a function of only $(t, \bar{X}(t), A)$, which is typically expressed as $\lambda^{G}(t \vert \bar{X}, A) = \lambda^{G}(t \vert \bar{X}(t), A)$. Furthermore, $\lambda^{G}$ is a hazard in the sense of defining the compensator of the censoring counting process $N^{C}(t) := \mathds{1}\{\tilde{\tau} \leq t, \Delta = 0\}$. More precisely, the processes
\begin{align*}
    M^{G}(t) &:= N^{C}(t) - \int_{0}^{t} \mathds{1}\{\Tilde{\tau} \geq u\}\lambda^{G}(u \vert \bar{X}(u), A) \diff u 
\end{align*}
is a censoring martingale with respect to the the measure $\prp_{\theta}$, and the filtration $\mathcal{F} := (\mathcal{F}(u))_{u \geq 0}$ generated by $\bar{X}, A$ and $\{C > u\}$ for each $u$. Under assumption of CAR, $M^{G}(t)$ is also a martingale with respect to $\prp_{\theta}$ and the censoring filtration $\tilde{\mathcal{F}} := (\tilde{\mathcal{F}}(t))_{t \geq 0}$, where 
\begin{align*}
\tilde{\mathcal{F}}(t) := \sigma(X(u) :  u \leq t, N^{C}(u) : u \leq t, A).
\end{align*}
Therefore, $M^{G}(t)$ depends only on information up to time $t$ \citep{Vaart2004}. It will be useful to distinguish the different terms of the influence function. To that end, let $IF_{0}$ be
\begin{align}
    IF_{0} := \frac{\mathds{1}\{A = a\}\Delta^{t_{0}} \mathds{1}\{Z(\tilde{\tau} \wedge t_{0} ) = j\}}{\pi(a\vert W) G(\tilde{\tau} \wedge t_{0}  \vert \bar{X}(C),A)} - \psi_{\obs}(\theta), \label{eq:IF_0}
\end{align}
where the first term is from the observed data estimand \eqref{eq:obs_estimand} and $\psi_{\obs}(\theta)$ is the target under $\prp_{\theta}$, which ensures that $IF_{0}$ has mean $0$ under $\prp_{\theta}$. The function $IF_{0}$ is then a valid score in the score space of $\mathbb{T}_{\theta}(P_{\theta})$, but is not orthogonalised with respect to the nuisance parameters $\pi$ and $G$, i.e., it is not orthogonal to $\mathbb{T}_{CAR}$. Therefore, applying step three will identify the efficient score. We denote the orthogonal projections of $IF_{0}$ onto $\mathbb{T}_{G}(P_{\theta})$ and $\mathbb{T}_{\pi}(P_{\theta})$ by $IF_{G}$ and $IF_{\pi}$, respectively. The resulting efficient influence function of $\psi_{\obs}$ at $P_{\theta}$ is:
\begin{align}
    IF_{\psi_{\obs}} := IF_{0} - IF_{\pi} - IF_{G}, \label{eq:Full_IF} 
\end{align}
where the projected terms can be represented as
\begin{align}
   IF_{\pi} &:= \left(\frac{\mathds{1}\{A=a\}}{\pi(a \vert W)} - 1 \right)Q(a,W), \text{ and } \\ 
   IF_{G} &:= - \frac{\mathds{1}\{A=a\}}{\pi(a \vert W)} \int_{0}^{\tilde{\tau} \wedge t_{0}}  \frac{ \eta(u,\bar{X}(u),a) }{G(u \vert \bar{X}(C), A)}\diff M^{G}(u),  \label{MGtermIF} 
\end{align}
with $\eta$ and $Q$ defined as
\begin{align}
    \eta(u,\bar{X}(u),a) &:= \e \left[ \frac{G(u \vert \bar{X}(C), A)}{G(\tilde{\tau} \wedge t_{0}  \vert \bar{X}(C), A)} \Delta^{t_{0}} \mathds{1}\{Z(\tilde{\tau} \wedge t_{0} ) = j\} \Big \vert \bar{X}(u), \tilde{\tau} \geq u, A=a \right], \label{eq:eta} 
\end{align}
for $u \leq t_{0}$, and 
\begin{align}
    Q(a,W) &:= \e\left[ \frac{\Delta^{t_{0}} \mathds{1}\{Z(\tilde{\tau} \wedge t_{0} ) = j\}}{G(\tilde{\tau} \wedge t_{0}  \vert \bar{X}(C), A)} \Big\vert A=a, W\right]. \label{eq:Q}
\end{align}

Since \eqref{eq:Full_IF} holds for each $\theta$ in our model, we can represent it through a parameter mapping of the target parameter $\psi_{\obs}(\theta)$ and the nuisance parameters, which we collectively denote $\gamma(\theta) := (\pi, G, \eta(G), Q(G))$, also emphasizing the dependence of $\eta$ and $Q$ on $G$. That is, for suitably chosen codomains $\Gamma$ of $\gamma(\cdot)$ and $\Psi$ of $\psi_{\obs}(\cdot)$ as functions of $\theta$, we can represent \eqref{eq:Full_IF} through a mapping $\varphi$ implicitly defined to satisfy
\begin{align}
    IF_{\psi_{\obs}}(\obs,\theta) &=:  \varphi(\obs, \gamma(\theta), \psi_{\obs}(\theta)). \label{eq:varphi}
\end{align}
We will allow a common misuse of notation, and denote the function $\varphi$ by $IF_{\psi_{\obs}}(\obs, \gamma(\theta), \psi_{\obs}(\theta))$. We allow the same for the different components of $IF_{\psi_{\obs}}(\obs,\theta)$. That is, we write $IF_{\pi}(\obs, \pi, Q)$ and $IF_{G}(O, G, \eta)$ for the parameter functions over appropriate domains, which we here leave unspecified, such that $IF_{\pi} =: IF_{\pi}(\obs, \pi, Q(G))$ and $IF_{G} =: IF_{\pi}(\obs, \pi, G, \eta(G))$. This representation, will be useful in Section \ref{sec:est_occ_prob}, when we define various estimators of our target parameter. Having derived the efficient influence function for state occupation probabilities, one can derive the efficient influence function of compact differentiable functionals of this parameters. An example would be the occupation probabilities integrated over a time intervals, so called expected length of stay (ELOS) which is discussed further in Appendix \ref{app:B:inf_fn}.

\section{Estimators of occupation probabilities}
\label{sec:est_occ_prob}

\subsection{Five proposed estimators}
\label{subsec:five}

Motivated by knowing the efficient influence function \eqref{eq:Full_IF} of $\psi_{\obs}$, we will now introduce five different estimators of occupation probabilities based on an iid sample $\obs_{1}, \ldots, \obs_{n}$ of size $n$. To that end, we introduce some common notation. We define the empirical measure $P_{n} := \frac{1}{n} \sum_{i=1}^{n} \delta_{\obs_{i}}$ and write $P_{n}f := \frac{1}{n} \sum_{i=1}^{n} f(\obs_{i})$ for integration functions $f$ of observed data values against the empirical measure. We will write $P_{\theta} f := \int f(o) \mathrm{d}P_{\theta}$ for the integration of $f$ with respect to an observed data distribution $P_{\theta}$. For functions $f(o, \gamma)$ of observed data and parameters we suppress notation of dependence on parameters and write $f = f(\cdot, \gamma)$ for a general $\gamma$ and $f_{0} = f(\cdot, \gamma_{0})$ for the true parameter $\gamma_{0}$. Furthermore, given a parameter estimator $\hat{\gamma}$ of $\gamma_{0}$ we will denote the plug-in as $\hat{f}(o) := f(o, \hat{\gamma})$. Observe that using this notation $P_{\theta}\hat{f}$, is a random object through the parameter estimator $\hat{\gamma}$. We motive the first three estimators as solutions to estimating equations in the target parameter using plug-in estimators $\hat{\pi},\hat{G}, \hat{Q}$ and $\hat{\eta}$ for the nuisance parameters. The inverse probability weighted (IPW) estimator is the solution in $\hat{\psi}$ to the estimating equation 
\begin{align*}
    P_{n} \hat{IF}_{0} = 0.
\end{align*}
That is, the IPW-estimator is
\begin{align}
\hat{\psi}_{n}^{0} &:= \frac{1}{n}\sum_{i=1}^{n}\left\{\frac{\mathds{1}\{A_{i} = a\}\Delta_{i}^{t_{0}} \mathds{1}\{Z_{i}(\tilde{\tau_{i}} \wedge t_{0} ) = j\}}{\hat{\pi}(a\vert W_{i}) \hat{G}(\tilde{\tau}_{i} \wedge t_{0} \vert \bar{X}_{i}(C_{i}),A_{i})} \right\}. \label{eq:IPW_est}
\end{align}
We emphasize that although we include the entire past up until censoring in the notation, the estimator $\hat{G}$ at any time point $u$ depends only on the history up until $u$, due to the coarsening at random assumption. In addition to the IPW estimator, we consider two, so-called augmented inverse probability weighted (AIPW) estimators, which we can likewise motive using $IF_{0} - IF_{\pi}$ respectively $IF_{0} - IF_{\pi} - IF_{G}$ to define estimating equations for the target parameter $\psi$. This leads to estimators
\begin{align}
    \hat{\psi}_{n}^{\nu} &:= P_{n} \hat{\nu}, \qquad \text{and}\label{eq:OS_nu_est} \\
    \hat{\psi}_{n}^{\mu} &:= P_{n} \hat{\mu}, \label{eq:OS_mu_est}
\end{align}
where 
\begin{align}
    \nu(\obs,\pi,G, Q) := \frac{\mathds{1}\{A = a\}\Delta^{t_{0}} \mathds{1}\{Z(\tilde{\tau} \wedge t_{0} ) = j\}}{\pi(a\vert W) G(\tilde{\tau} \wedge t_{0}  \vert \bar{X}(C),A)} - IF_{\pi}(\obs, \pi,Q), \label{eq:nu}
\end{align}
and
\begin{align}
    \mu(\obs,\pi,G, Q, \eta) := \nu(\obs,\pi,G, Q) -  IF_{G}(\obs, G,\eta). \label{eq:mu}
\end{align}
$\hat{\psi}_{n}^{\mu}$ is also the one-step estimator based on the estimating function $IF_{\psi_{\obs}}$, since
\begin{align*}
    \hat{\psi}_{n}^{\mu} &= \hat{\psi}_{n}^{0} + P_{n}IF_{\psi_{\obs}}(\hat{\gamma}, \hat{\psi}_{n}^{0}).
\end{align*}
The last two estimators are so-called modified AIPW estimators, modifying $\hat{\psi}_{n}^{\nu}$ and $\hat{\psi}_{n}^{\mu}$, respectively. Define the modified $\mu$ estimator as
\begin{align}
    \hat{\psi}_{n}^{\mu, mod} := P_n \left(\frac{\mathds{1}\{A = a\}\Delta^{t_{0}} \mathds{1}\{Z(\tilde{\tau} \wedge t_{0} ) = j\}}{\hat \pi(a\vert W) \hat G(\tilde{\tau} \wedge t_{0}  \vert \bar{X}(C),A)} - \hat \beta_{\pi} IF_{\pi}(\obs, \hat \pi,\hat Q) - \hat \beta_G IF_{G}(\obs, \hat G,\hat \eta)\right), \label{eq:mu_mod} 
\end{align}
where the coefficients $\hat \beta_{\pi}$ and $\hat \beta_G$ are estimated by OLS regression. We regress $\hat{IF}_0$ on the estimated projection term components $-IF_{\pi}(\obs, \hat \pi,\hat Q)$ and $-IF_{G}(\obs, \hat G,\hat \eta)$, with no intercept. The intuition behind this is to force orthogonality of the terms, when it is violated due to biased nuisance estimators. This may have beneficial properties in efficiency over the unmodified AIPW estimators when the estimators for $\hat \eta$ and $\hat Q$ are biased. The modified $\nu$ estimator is defined analogously as
\begin{align}
    \hat{\psi}_{n}^{\nu, mod} := P_n \left(\frac{\mathds{1}\{A = a\}\Delta^{t_{0}} \mathds{1}\{Z(\tilde{\tau} \wedge t_{0} ) = j\}}{\hat \pi(a\vert W) \hat G(\tilde{\tau} \wedge t_{0}  \vert \bar{X}(C),A)} - \hat \beta_{\pi} IF_{\pi}(\obs, \hat \pi,\hat Q)\right). \label{eq:nu_mod} 
\end{align}
For more details about the efficiency gain, see Section \ref{subsec:prop} and \citet{LR2003}[Section 2.5-2.7].

\newpage

\subsection{Nuisance parameter estimators}
\label{subsec:nuisance}

\centerline{\textit{$\pi$ and $G$: Working models and MLE}}

To estimate $\pi_{0}$ and $G_{0}$, we specify \textit{working models}, i.e. models which may or may not be true. For $\pi_{0}$ we assume a logistic regression model and for $G_{0}$ we assume a Cox model. Under such parameterization, the tangent spaces are altered and likewise the influence function of the estimand \eqref{eq:obs_estimand} defining the efficiency bound. To see how, in the simple survival setting see e.g., \cite{Hubbard2000} and \cite{LHR2002}. Even if true, we will not rely on such parametric constraints, rather we stick to the estimators suggested by the parameterization only restricting our model by the stated identifiable criteria. Consequently, if the working models are true, the One-step estimator \eqref{eq:OS_mu_est} is not efficient. In our implementation, $\hat{\pi}$ is the plug-in estimator of $\pi_{0}$ based on the maximum likelihood estimator of the regression coefficients in a logistic model. The estimator of $G_{0}$ is constructed as follows: Let $\hat{\Lambda}_{0}$ be the Breslow estimator of the baseline cumulative censoring hazard (see e.g. \cite{Lin2007}). That is, 
\begin{align*}
    \hat{\Lambda}_{0}(t) :=  \int_{0}^{t} \frac{\mathrm{d}\bar{N}^{C}(u)}{\sum_{i=1}^{n} \mathds{1}\{\tilde{\tau_{i}} \geq u\} \exp(\hat{\beta}^{T}\tilde{X}_{i}(u))}
\end{align*}
where $\bar{N}^{C} := \sum_{i=1}^{n} N_{i}^{C}$, $\tilde{X}_{i}$ is a vector of possibly time dependent covariates, predictable with respect to the censoring filtration, and $\hat{\beta}$ is the Cox partial likelihood estimator of the regression coefficients. The fitted value of the censoring survival function for an individual with at risk process $\mathds{1}\{\tilde{\tau} \geq u\}$ and covariate process $\tilde{X}$ is then the product integral
\begin{align*}
    \hat{G}_{n}(t \vert \bar{X},A) := \prod_{(0,t]}\left(1 - \mathrm{d}\hat{\Lambda}_{n} \right),
\end{align*}
where $\hat{\Lambda}_{n}(t) := \int_{0}^{t} \mathds{1}\{\tilde{\tau} \geq u\} \exp(\hat{\beta}^{T}\tilde{X}(u))\mathrm{d}\hat{\Lambda}_{0}(u)$. \\

\centerline{\textit{$Q$ and $\eta$: The regression approach}}
\label{subsec:Q_eta_regapproach}

To estimate the nuisance parameters $(Q_{0}, \eta_{0})$ we regress the weighted observations, sometimes called pseudo-outcomes (see e.g. \cite{Kennedy23}), against summaries of the conditioning set. This idea was introduced by \cite{FanGjib1994}, and has been used extensively to estimate parameters defined by latent time to event data and other missing data structures; see e.g. \cite{Robins1992}, \cite{Hubbard2000}, \cite{LR2003}, \cite{RubinLaan2007}. More precisely, we estimate $\eta$ based on estimates $\hat{G}$, by regressing within each treatment group $A = a$, the weighted observations
\begin{align*}
    \obs_{\hat{G}}(u,a) := \frac{\mathds{1}\{Z(\tilde{\tau} \wedge t_{0} ) = j\}\Delta^{t_{0}}\hat{G}(u \vert \bar{X},a)}{\hat{G}(\tilde{\tau} \wedge t_{0}  \vert \bar{X},a)}
\end{align*}
against summaries of the data $\bar{X}(u), \tilde{\tau} \wedge t_{0}  \geq u, A=a$ over a specified time-grid of $u$'s. Likewise, $Q$ is estimated by regressing, within each treatment group $A = a$, $\frac{\mathds{1}\{Z(\tilde{\tau} \wedge t_{0} ) = j\} \Delta^{t_{0}}}{\hat{G}(\tilde{\tau} \wedge t_{0}  \vert \bar{X},a)}$ against $W$. Many estimation procedures suitable for estimating conditional expectations could be applied. \cite{FanGjib1994} uses local polynomial regression, and in \cite{Hubbard2000} generalized additive models are used. In the simulation experiment we use a simple second order polynomial regression. See Section \ref{sec:4:simulations} for details.

\newpage

\subsection{Properties of estimators}
\label{subsec:prop}
\centerline{\textit{Robustness and efficiency}}

We now discuss the large sample properties of the five estimators. More precisely, we discuss consistency and efficiency of the target parameter estimators, and robustness to possibly misspecified nuisance parameter estimators. The arguments for robustness made in this section are suggestive and based on super-population bias properties of the parameter mappings defining the estimators rather than proofs of consistency. 

Let $\gamma_{0} = (\pi_{0}, G_{0}, Q_{0}, \eta_{0})$ be the nuisance parameter values under the true data generating distribution $P_{\theta_{0}}$. Let $\gamma_{1} = (\pi_{1}, G_{1}, Q_{1}, \eta_{1})$ be other nuisance parameter values, which we will interpret as the limit (appropriately defined) in probability for the nuisance parameter estimators. We argue for robustness properties of $\hat{\psi}_{n}^{\mu}$, by conditions on $\gamma_{1}$, such that
\begin{align}
    E_{\theta_{0}}[\mu(\obs,\gamma_{1})] &= E_{\theta_{0}}[\mu(\obs,\gamma_{0})]. \label{robust_mu}
\end{align}
In Appendix \eqref{app:C:Robust_and_mod_est}, we find that this is true if ($\eta_{1} = \eta_{0}$ or $G_{1} = G_{0})$ and ($\pi_{1} = \pi_{0}$ or $Q_{1} = Q_{0})$ is true. However, considering a general $\eta_{1}$ and $Q_{1}$ is somewhat artificial, since both our identification formulae insist that $\eta$ and $Q$ depend on $G$ and the regression approach for estimating $Q_{0}$ and $\eta_{0}$ also results in dependencies of $\hat{\eta}$ and $\hat{Q}$ on $\hat{G}$. Therefore, a more appropriate statement is that \eqref{robust_mu} is true if $G_{1} = G_{0}$ and ($\pi_{1} = \pi_{0}$ or $Q_{1} = Q_{0})$ is true. The parameter mapping $\nu$ in \eqref{eq:nu} is unbiased under the same conditions suggesting the same robustness properties for $\hat{\psi}_{n}^{\nu}$. Similar considerations suggest that the IPW estimator is consistent if $G_{1} = G_{0}$ and $\pi_{1} = \pi_{0}$ and likewise for the modified estimators. To explain the hierarchy of efficiency, let $\Vert \cdot \Vert$ denote the $L_{2}(P_{\theta_{0}})$-norm, let $M$ be some non-trivial closed subspace of the tangent space and denote by $proj_{M}$ the orthogonal projection onto $M$. By orthogonal decomposition
\begin{align*}
        \Vert IF_{0} \Vert &= \Vert proj_{M}(IF_{0}) \Vert + \Vert IF_{0} - proj_{M}(IF_{0}) \Vert \geq  \Vert IF_{0} - proj_{M}(IF_{0}) \Vert
\end{align*}
In particular, this is true for $M = \mathbb{T}_{CAR}$ or M defined as the closed span by $IF_{\pi}$. To see the efficiency gain of modified estimators denote $\tilde{IF}_{\pi} := IF(\obs, \pi_{0},Q_{1})$ and $\tilde{IF}_{G} = IF_{G}(O, \pi_{0}, G_{0}, \eta_{1})$. These vectors are orthogonal, thus forming an un-normalized basis for their closed span, call it $M$. It then follows, by definition of orthogonal projections, that for any $U \in M$, then
    \begin{align*}
        \Vert IF_{0} - U\Vert \geq \Vert IF_{0} - proj_{M}(U)\Vert  
    \end{align*}
    where the projection is given by
    \begin{align*}
        proj_{M}(U) &= \frac{\langle U, \tilde{IF}_{\pi} \rangle}{\Vert \tilde{IF}_{\pi} \Vert^{2}} \tilde{IF}_{\pi} + \frac{\langle U, \tilde{IF}_{G} \rangle}{\Vert \tilde{IF}_{G} \Vert^{2}} \tilde{IF}_{G}.
    \end{align*}
In particular, the distance is minimized for $U = proj_{M}(IF_{0})$, which is the super-population argument for improved efficiency by the modified estimators.  \\

\centerline{\textit{Convergence rates}}

A desirable feature of the one step-estimator, and more generally targeted learning procedures, is the enabling of machine learning estimators to estimate nuisance parameters while ensuring $\sqrt{n}$ consistent target parameter estimators, which is crucial for subsequent inference. In this section we discuss the convergence rate of the one step estimator. In doing so we will use small and big stochastic order notation $o_{p}(\cdot)$ and $O_{p}(\cdot)$ (see \cite{vaart_1998} for a definition). To gain insight into the convergence rates of the one-step estimator, we first consider a more general estimation problem. Given a random variable $\obs \sim P_{0}$ suppose we define a target parameter $\psi_{0} := P_{0} f_{0}(\gamma_{0})$ for a function $f(o, \gamma)$ depending on observations through $o$ and nuisance parameters $\gamma$. Then we have the following decomposition 
\begin{align*}
    P_{n} \hat{f} - P_{0}f_{0} &= (P_{n} - P_{0})f_{0} + \underbrace{(P_{n} - P_{0})(\hat{f} - f_{0})}_{Emp(\hat{f},f_{0})} + \underbrace{P_{0}(\hat{f} - f_{0})}_{R(\hat{f},f_{0})}
\end{align*}
If it holds that $Emp(\hat{f},f_{0}) = o_{p}(1/\sqrt{n})$ and $R(\hat{f},f_{0}) = o_{p}(1/\sqrt{n})$
then 
\begin{align}
    P_{n} \hat{f} - P_{0}f_{0} &= P_{n}\{f_{0} - \psi_{0}) + o_{p}(1/\sqrt{n}) \label{Assymp_linear}
\end{align}
i.e. $f_{0} - \psi_{0}$ is the influence function of the estimator $\hat{\psi}_{n} := P_{n} \hat{f}$. The term $Emp(\hat{f},f_{0})$ is typically called the empirical process term (hence the notation) and $R(\hat{f},f_{0})$ the remainder term. Indeed, it is simple to verify that $R(\mu(\cdot, \gamma(\theta)), \mu_{0}) = R(P_{\theta}, P_{\theta_{0}})$ where $R(P_{\theta}, P_{\theta_{0}})$, is the Von-Mises remainder term for $\psi_{O}(P_{\theta})$ defined as:
\begin{align*}
    R(P_{\theta}, P_{\theta_{0}}) &:= \psi_{O}(P_{\theta}) - \psi_{O}(P_{\theta_{0}}) + E_{\theta_{0}}\left[ IF(O, \gamma(\theta)) \right].
\end{align*}
In appendix \eqref{app:D:rates} we argue that
\begin{align}
    R(\hat{\mu}, \mu_{0}) &= P_{0}\left\{ \left(\frac{\pi_{0}}{\hat{\pi}} - 1 \right) \left( Q_{0}(G_{0}) - \hat{Q}(\hat{G}) \right) \right\} \ + \label{remainder} \\
    &P_{0}\left\{ \left(\frac{\pi_{0}}{\hat{\pi}} - 1 \right) \int\frac{(\hat{\eta}(\hat{G}) - \eta_{0}(G_{0})}{\hat{G}} \mathrm{d}(\Lambda^{G_{0}} -\hat{\Lambda}^{G}) \right\} \ + \nonumber 
 \\
    &P_{0}\left\{\int\frac{(\hat{\eta}(\hat{G}) - \eta_{0}(G_{0})}{\hat{G}} \mathrm{d}(\Lambda^{G_{0}} -\hat{\Lambda}^{G}) \right\} \nonumber 
\end{align}
and that under the assumptions below $\hat{\psi}^{\mu}$ satisfies \eqref{Assymp_linear} i.e. it is asymptotically linear with influence function
$IF_{\psi_{O}}$. \\

\noindent \textbf{Assumptions:}
\begin{enumerate}
    \item Positivity assumptions: There exist positive constants $\delta^{\hat{\pi}}$ and $\delta^{\hat{G}}$ such that for all $n$,
    $\hat{\pi} \geq \delta^{\hat{\pi}}$ almost surely and $\hat{G} \geq \delta^{\hat{G}}$ almost surely.
    \item Rate assumptions: Let $r_{n}^{\hat{\pi}}, r_{n}^{\hat{G}}$ and $r_{n}^{\hat{Q}}$ be sequences of real numbers such that the reciprocal goes to zero as $n \to \infty$. Assume the following rate conditions holds for the nuisance parameter estimators $\Vert \pi_{0} - \hat{\pi}  \Vert = O_{p}(1/r_{n}^{\hat{\pi}})$, $\Vert Q_{0}(G_{0}) - \hat{Q}(G_{0}) \Vert = O_{p}(1/r_{n}^{\hat{Q}})$, $\Vert \hat{G} - G_{0} \Vert = O_{p}(1/r_{n}^{\hat{G}})$ and $\Vert \hat{Q}(G_{0}) - \hat{Q}(\hat{G}) \Vert = O_{P}(1/r_{n}^{\hat{G}})$. Assume the following key constraints on the rates
    \begin{align}
        \sqrt{n}\left( \frac{1}{r_{n}^{\hat{\pi}} r_{n}^{\hat{G}}} + \frac{1}{r_{n}^{\hat{\pi}}r_{n}^{\hat{Q}}}  \right) &= o(1) \label{product_rate_condition} \\
    \left\Vert \int (\hat{\eta}(\hat{G}) - \eta_{0}(G_{0}) \mathrm{d}(\Lambda^{G_{0}} -\hat{\Lambda}^{\hat{G}})  \right\Vert &= o_{p}(1/\sqrt{n}) \label{MG_assumption}
    \end{align}
    \item Empirical process assumptions: Assume that  $\Vert \hat{\mu} - \mu_{0} \Vert = o_{p}(1)$ and that $\hat{\mu}$ takes values in a Donsker class.
\end{enumerate}

The condition that $\Vert \hat{Q}(G_{0}) - \hat{Q}(\hat{G}) \Vert = O_{P}(1/r^{\hat{G}})$ will in many cases be satisfied if $\hat{Q}$ is a linear smoother. That is, as a function of baseline values $w$ in the support of $W$, we have
\begin{align*}
    \hat{Q}(G)(w) &= \sum_{i:A_{i} = a} S_{i,N_{a}}(w)\frac{\mathds{1}\{Z_{i}(\tilde{\tau_{i}} \wedge t_{0} ) = j\} \Delta_{i}^{t_{0}}}{G(u \vert X_{i},a)}
\end{align*}
where $S_{i,N_{a}}(w)$ are individual specific weights  depending on all $N_{a}$ baseline variables $W_{j}$ for which $A_{j} = a$, and is determined by the choice of smoother. Many linear smoothers will satisfy the conditions of Stones theorem (see e.g. \cite{Gyorfi2002} Theorem 4.1). In particular,
\begin{align*}
   \lim_{N_{a} \to \infty} E\left[\sum_{i: A_{i} = a}S_{i,N_{a}}(W)^{2} \right] = 0.
\end{align*}
By Cauchy-Swartz for finite vectors we have
\begin{align*}
    &\vert \hat{Q}(G_{0})(w) - \hat{Q}(\hat{G})(w) \vert \leq \\
    &\left\{ \sum_{i:A_{i} = a} S_{i,N_{a}}(w)^{2} \right\}^{\frac{1}{2}} \left\{ \sum_{i:A_{i} = a} \left( \frac{\mathds{1}\{Z_{i}(\tilde{\tau_{i}} \wedge t_{0} ) = j\} \Delta_{i}^{t_{0}}}{G_{0}(u \vert X_{i},a)} - \frac{\mathds{1}\{Z_{i}(\tilde{\tau_{i}} \wedge t_{0} ) = j\} \Delta_{i}^{t_{0}}}{\hat{G}(u \vert X_{i},a)} \right)^{2} \right\}^{\frac{1}{2}}
\end{align*}
Therefore, 
\begin{align*}
    &\Vert \hat{Q}(G_{0}) - \hat{Q}(\hat{G}) \Vert^{2} \leq \\
    & E\left[  \sum_{i:A_{i} = a} S_{i,N_{a}}(W)^{2} \right]  \sum_{i:A_{i} = a} \left( \frac{\mathds{1}\{Z_{i}(\tilde{\tau_{i}} \wedge t_{0} ) = j\} \Delta_{i}^{t_{0}}}{G_{0}(u \vert X_{i},a)} - \frac{\mathds{1}\{Z_{i}(\tilde{\tau_{i}} \wedge t_{0} ) = j\} \Delta_{i}^{t_{0}}}{\hat{G}(u \vert X_{i},a)} \right)^{2} 
\end{align*}
The first term is $O_{p}(1)$ and if the second term is $O_{p}(1 / (r_{n}^{\hat{G}})^{2})$, then 
$\Vert \hat{Q}(G_{0}) - \hat{Q}(\hat{G}) \Vert^{2} = O(1/r_{n}^{\hat{G}})$. Condition 3. is to ensure $Emp(\hat{\mu},\mu_{0}) = o_{p}(1/\sqrt{n})$ and the Donsker class condition can be relaxed by instead using data-splitting and cross fitting as discussed e.g. in \cite{kennedy2023semiparametric}. The assumption \eqref{MG_assumption} is clearly problematic in that we would want conditions under which this property is satisfied. As discussed in \cite{Munch2023} p.24 it is not obvious, even in the simplest cases, how to select appropriate norms on the integrand and the integrator such that product rate condition similar to \eqref{product_rate_condition} can be established to ensure \eqref{MG_assumption}. This is an open problem, which we do not address in this paper. Using the Breslow estimator, $r_{n}^{\hat{G}} = \sqrt{n}$ has been established in simpler cases when censoring dependents on at most bounded baseline covariates (see \cite{GerdsScheike2008} and \cite{Nane2013}), but it is expected to hold more generally for time dependent covariates under the conditions of \cite{GillAndersen1982}. If $r_{n}^{\hat{G}} = \sqrt{n}$, the requirement \eqref{product_rate_condition}, becomes 
\begin{align*}
    \frac{\sqrt{n}}{r_{n}^{\hat{\pi}}r_{n}^{\hat{Q}}} &= o(1). 
\end{align*}
Let $d$ be the fixed dimension of the baseline covariates $W$ and let $\pi$ and $Q$, as functions of their conditioning argument, be functions of sub-vectors of $W$ of dimension $d_{\pi} \leq d$ and $d_{Q} \leq d$. If $\pi$ and $Q$ are $\alpha$-smooth respectively $\beta$-smooth and $\hat{\pi}$ and $\hat{Q}$ are minimax optimal, then \eqref{product_rate_condition} reduces to the standard requirement that 
\begin{align*}
    \frac{d_{\pi} d_{Q}}{4} < \alpha \beta.
\end{align*}

\newpage

\section{Simulation testing and results}
\label{sec:4:simulations}

To assess the performance and illustrate properties of the estimators discussed in Subsection \ref{subsec:five}, we conduct various simulation experiments. In Subsection \ref{subsec:scenario_setup}, we set up an example framework, extract the true  occupation probabilities $\psi_0(t)$ from the generated latent data and censor them to obtain "observed" data. We implemented the five estimators described in Section \ref{subsec:five} in R, and ran the implementations on these simulated observational datasets. The IPW estimator $\hat{\psi}_{n}^{0}$ is used as a simple benchmark method to analyse the properties of our two proposed AIPW estimators $\hat{\psi}_{n}^{\nu}$ and $\hat{\psi}_{n}^{\mu}$, and the corresponding modified AIPW estimators $\hat{\psi}_{n}^{\nu, mod}$ and $\hat{\psi}_{n}^{\mu, mod}$. The five estimators are run under various conditions, where either all nuisance plug-in estimators are consistent (Subsection \ref{subsec:EffandCvg}), or the estimators for one of $\pi_0$, $G_{0}$, $\eta_0$, or $Q_0$ are biased (Subsection \ref{subsec:misspecified}). We bias the nuisance estimators by misspecifying the formulas, either by removing variables, or using incorrect functional forms of the correct variables. 

\subsection{Example Scenario Set-up}
\label{subsec:scenario_setup}

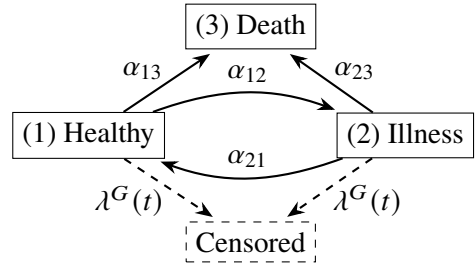
\begin{figure}[H]
    \centering
\begin{tikzpicture}[
    array/.style={rectangle split, 
        rectangle split parts = 5,
        rectangle split horizontal,
    minimum height = 2em
    }, baseline=0pt
]
 \node[draw] (1) {$\text{(1) Healthy}$};
  \node [right=of 1] (2) {};
   \node[draw] [right=of 2] (3) {$\text{(2) Illness}$}; 
   \node[draw] [above=of 2] (4) {$\text{(3) Death}$};
   \node[draw, dashed] [below=of 2] (5) {$\text{Censored}$};
 \draw[-Stealth, thick, shorten >=3pt] (1) to [out=20, in=160] node[above, pos=0.50] {$\alpha_{12}$} (3);       
 \draw[-Stealth, thick, shorten >=3pt] (3) to [out=-160, in=-20] node[above, pos=0.50] {$\alpha_{21}$} (1);     
 \draw[-Stealth, thick, shorten >=3pt] (1) to node[above, pos=0.35, xshift = -5pt] {$\alpha_{13}$}(4);     
 \draw[-Stealth, thick, shorten >=3pt] (3) to node[above, pos=0.35, xshift = 5pt] {$\alpha_{23}$} (4);   
 \draw[-Stealth, thick, dashed, shorten >=3pt] (1) to node[below, pos=0.25, xshift = -7pt] {$\lambda^G(t)$}(5);     
 \draw[-Stealth, thick, dashed, shorten >=3pt] (3) to node[below, pos=0.25, xshift = 7pt] {$\lambda^G(t)$} (5);   
\end{tikzpicture}
\caption{The illness-death model with recurrence and censoring. In our example, transition rates between states $\alpha_{ij} := \alpha_{ij}(W,A) \in \mathbb{R}^+$ depend only on the individuals' baseline covariates, and are constant in time. Censoring rates $\lambda^G(t)$ are not constant, because they depend on the individuals' transition histories up to each time $t$, as defined in \eqref{eq:Ghaz_def}, which we model by \eqref{eq:G_correct_form}.}
\label{fig:multistate_model_dag}
\end{figure}

We aim to simulate a dataset that aligns with plausible real world data dynamics. This grants us a more accurate test of efficiency and robustness than using a more simplified toy dataset. We define a simple model of a three state illness-death process with recurrence, where we assume that all patients start from the \textit{Healthy} state, as illustrated in Figure \ref{fig:multistate_model_dag}.

\begin{figure}[H]
    \centering
    \includegraphics[width=\linewidth]{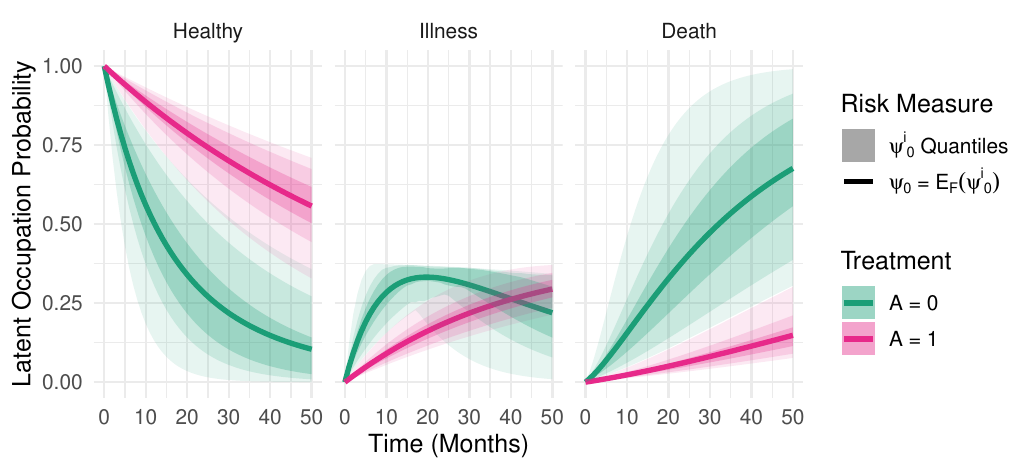}
    \caption{Latent occupation probability curves $\psi_0(t|a)$, given treatment $A=a$, for the population average and bands between the minimum, the quantiles $10\%$, $25\%$, $75\%$, $90\%$, and the maximum values of $\psi_0^i(t|a)$.}
    \label{fig:true_occprob}
    \end{figure}

The baseline variables $(W_i)_{i=1,2,3}$ are chosen to mimic observations of \textit{Sex}, \textit{Age} and \textit{BMI} in a typical population. We use sigmoid transformations $f_1(\textit{Age})$ of \textit{Age} and $f_2(\textit{BMI})$ of \textit{BMI} to capture non-linear impact of the variables on risk. This allows us to capture high risk in both extremes of the \textit{BMI} scale, for example. See Figure \ref{fig:BMI_Age_transf} in Appendix \ref{app:E:simresults} for the exact transformations $f_1$ and $f_2$. The latent data process $Z_A(t)$ is simulated as a continuous-time Markov chain (CTMC) dependent on the baseline variables, including treatment. We simulated a setting where the hypothetical treatment is beneficial against a hypothetical disease, but the occupation probabilities \eqref{eq:target} of both the treated and non-treated populations are mostly still bounded from 0\% and 100\% within $(0,50]$. We ensure that this holds for even the most extreme covariate value combinations, as demonstrated in Figure \ref{fig:true_occprob}. The treatment assignment of each individual is simulated by calculating the propensity scores with a logistic model, conditional on their baseline covariates:
    \begin{align}
        & A|W =w \sim \text{Bernoulli}(\pi_0(1|w)),  \text{ where} \notag \\ 
        & \pi_0(1|w) := \text{expit}\left(\xi_0 + \xi_1 f_1(\textit{Age}) + \xi_2 f_2 (\textit{BMI}) \right). \label{eq:pi_correct_form}
    \end{align}
We right-censor the latent data using censoring times simulated through a Cox model with a constant baseline hazard $\beta_0(t) :\equiv \beta_0 \in \mathbb{R}$. We use baseline variables, but also include time-dependent summary variables of the latent data process to model dependence on $\bar{Z}_A(t-)$. Namely, we use an indicator of the current state value being \textit{Illness}, $\mathds{1}\{Z_A(t) = 2\}$, and the first time the individual transitioned to \textit{Healthy}. For the latter variable, we first define the actual stopping time $\tau_{1\to2} := \inf\{t \geq 0: Z_A(t)=2\}$ and use it to define the process $\tau_{12}(t) :=  \tau_{1\to2} \cdot \mathds{1}\{ \tau_{1 \to 2} \leq t\}$. The $\mathcal{F}_{t-}$ measurable censoring hazard at time $t$ is then
\begin{align}
    \lambda^{G_0}(t|\mathcal{F}_{t-}) := \beta_0 \exp ( \overbrace{\beta_1 A + \beta_2 f_1(\textit{Age}) + \beta_3 f_2(\textit{BMI})}^{\text{baseline variables}} + \overbrace{\beta_4 \mathds{1}\{Z_A(t-) = 2\} + \beta_5 \tau_{12}(t-)}^{\text{history up to time } t-})), \label{eq:G_correct_form}
\end{align}
which is used to construct a discretized cumulative hazard grid for simulation purposes.

In order to benchmark our estimators' performances to the true occupation probability curves $\psi_0(t)$ of the model, as defined in \eqref{eq:target}, we must ensure that we are able to first approximate the latter with minimal variance. We could use Monte Carlo (MC) approximation on the simulated paths, but this may converge slowly when individuals have significant heterogeneity. Instead, simulating $Z_A$ as a baseline dependent CTMC allows an initial closed form calculation of the conditional occupation probability curves for an individual. This removes a layer of uncertainty, so MC approximation with these conditional probability curves results in a lower standard error than on the realised paths. We determined that a population of $50000$ was sufficient for MC convergence. The closed form value for individual $i$ is $\psi^{i}_0(t|a) := (1,0,0)\cdot\exp(Q_i(a)t)$, where $Q_i(a)$ is the infinitesimal generator using the constant transition rates $\alpha(w_i,a)$ defined by their baseline covariates $w_i$ and treatment set to $0$ or $1$. 

Using the framework introduced above, we simulate $100$ independent datasets with $9000$ individuals' observed state transition histories in each.  All simulations and calculations were performed in R. In the simulations, around $25\%$ of individuals end up censored, as shown in Figure \ref{fig:observed_stackplot} in Appendix \ref{app:E:subsec:scendesign}. Further specifications of the simulation, such as the risk transformations $f_1, f_2$ and the numeric values of the vectors $\beta$, $\xi$ and the CTMC coefficient matrix are also detailed in Appendix \ref{app:E:subsec:scendesign}. 

\subsection{Testing Bias, Efficiency and Coverage}
\label{subsec:EffandCvg}
To test the performance of the implemented estimators $\hat{\psi}_{n}^{0}$, $\hat{\psi}_{n}^{\nu}$, $\hat{\psi}_{n}^{\mu}$, $\hat{\psi}_{n}^{\nu, mod}$ and $\hat{\psi}_{n}^{\mu, mod}$, we run them with consistent plug-in estimators of the true nuisance parameters. 
\begin{itemize}
    \item For $\pi_0$ and $\lambda^{G_0}(t)$ we define the consistent estimators $\hat \pi_0$ and $\hat \lambda^G_0(t)$ by specifying the correct variables and functional forms from the true models \eqref{eq:pi_correct_form} and \eqref{eq:G_correct_form} and estimate the coefficients using logistic regression and Cox regression, respectively, as discussed in Section \ref{subsec:nuisance}.
    \item For $\eta_0(t,u)$ and $Q_0(u)$, there is no closed form solution for the true functional form. Instead, we approximate the true functional forms using a linear formula, allowing for first and second order covariate terms where applicable. As discussed in Section \ref{subsec:Q_eta_regapproach}, we then run a linear regression on weighted observations. For simplicity, we provide these formulas in R syntax, where the \textit{Poly}(2) function from the \texttt{stats} package creates second order orthogonal polynomials of the variables, and the "*" operator creates both individual and interaction terms. The estimators for these nuisance parameters are then
\end{itemize}
\begin{align}
    \hat Q_0(u) & \sim A * (\text{Sex} + \mathrm{Poly}(f_1(\text{Age}),2)  + \mathrm{Poly}(f_2(\text{BMI}),2)), \text{ and } \\
    \hat \eta_0(t,u) & \sim A + \text{Sex} + \mathrm{Poly}(f_1(\text{Age}),2)  + \mathrm{Poly}(f_2(\text{BMI}),2) \mathds{1}\{Z(t-) = 2\} + \tau_{12}(t-).
\end{align}

\begin{figure}
    \centering
    \includegraphics[width=\linewidth]{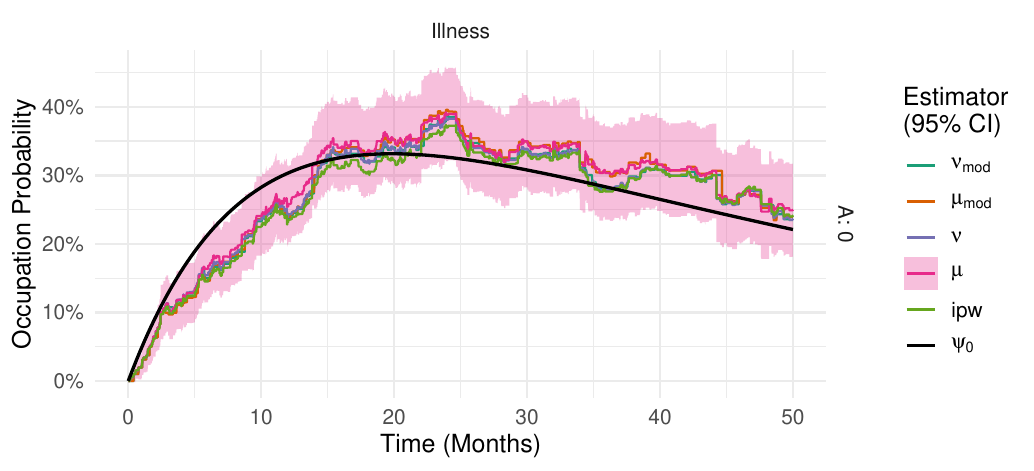}
    \caption{Example run of the estimators $\hat{\psi}_{n}^{0}$, $\hat{\psi}_{n}^{\nu}$, $\hat{\psi}_{n}^{\mu}$, $\hat{\psi}_{n}^{\nu, mod}$ and $\hat{\psi}_{n}^{\mu, mod}$, all using the consistent nuisance parameter estimators $\hat \pi_0$, $\hat \lambda^{G}_0$, $\hat Q_0$ and $\hat \eta_0$. For $\hat{\psi}_{n}^{\mu}$, we include the $95$ \% pointwise confidence intervals. Population size is $500$, on a single example dataset.}
    \label{fig:correctly_spec_examplerun}
\end{figure}

 The fitted influence functions are used to calculate $95\%$ point-in-time confidence intervals for the estimators $\hat{\psi}_{n}^{\mu}$ and $\hat{\psi}_{n}^{\mu, mod}$ using these standard deviations and Gaussian approximation. As an example of the estimation code output, see Figure \ref{fig:correctly_spec_examplerun}, where we present the output of the estimators on a dataset of $500$ observations, with the $95\%$ confidence interval depicted for $\hat{\psi}_{n}^{\mu}$. 
We present a discussion of the results below.

    \newpage
    \begin{figure}
        \centering
        \includegraphics[width=\linewidth]{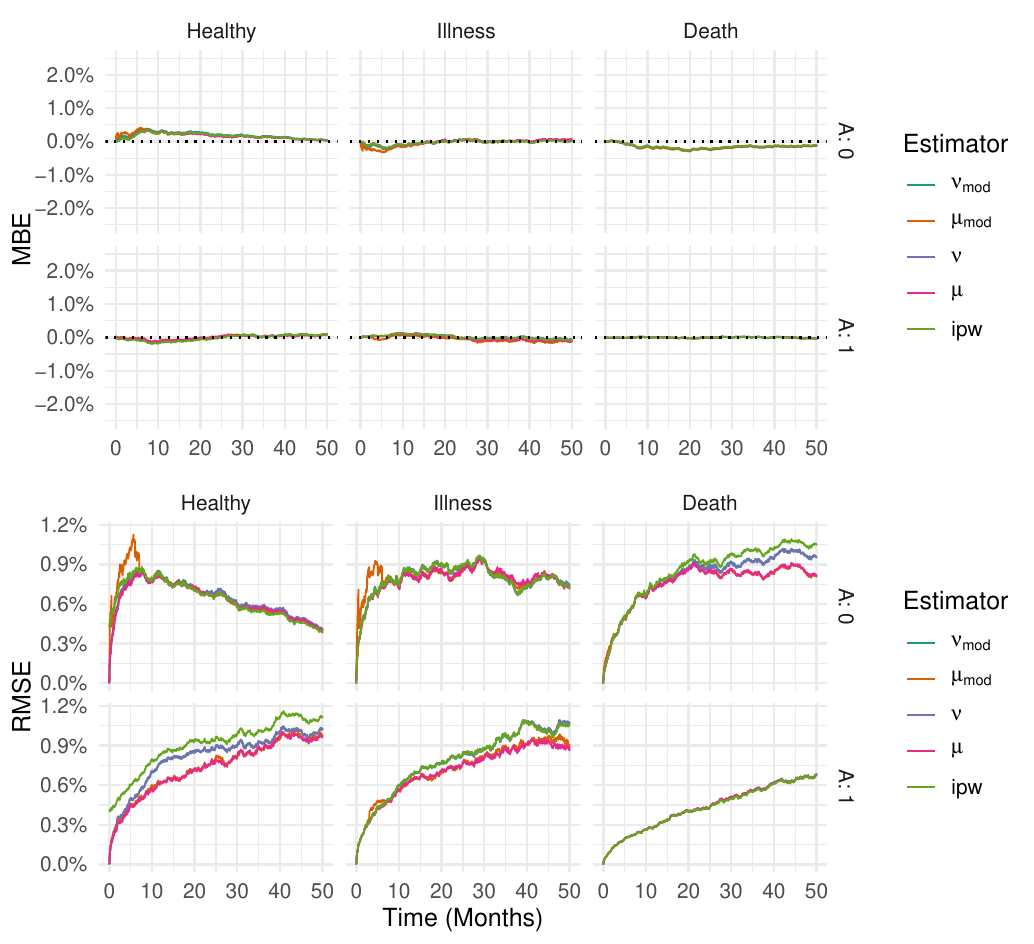}
        \caption{MBE and RMSE metrics of the estimators $\hat{\psi}_{n}^{0}$, $\hat{\psi}_{n}^{\nu}$, $\hat{\psi}_{n}^{\mu}$, $\hat{\psi}_{n}^{\nu, mod}$ and $\hat{\psi}_{n}^{\mu, mod}$, all using the consistent nuisance parameter estimators $\hat \pi_0$, $\hat \lambda^{G}_0$, $\hat Q_0$ and $\hat \eta_0$. Averages are over $100$ independent datasets with a population size of $9000$ in each.}
        \label{fig:Test1_n9000_MBE_RMSE}
    \end{figure}
 The resulting outputs are compared to the true occupation probability curves $\psi_0(t|a)$ from Figure \ref{fig:true_occprob} to examine the mean bias error (MBE) and root mean square errors (RMSE). We expect all estimators to be unbiased, with $\hat{\psi}_{n}^{\nu}$ having less RMSE than $\hat{\psi}_{n}^{0}$, and $\hat{\psi}_{n}^{\mu}$ having the least. The MBE and RMSE results are presented in Figure \ref{fig:Test1_n9000_MBE_RMSE}, and indeed, the MBE is $\approx 0$ for all estimators. In terms of efficiency, we can see that the estimator $\hat{\psi}_{n}^{\nu}$ converges with slightly less RMSE than $\hat{\psi}_{n}^{0}$ in two plots, but overlaps with $\hat{\psi}_{n}^{0}$ in the \textit{Illness} state plots. The estimator $\hat{\psi}_{n}^{\mu}$, however, is has less RMSE than both, with a visible gap in $3$ out of $6$ plots.

    \newpage 
    \begin{figure}
        \centering
        \includegraphics[width=\linewidth]{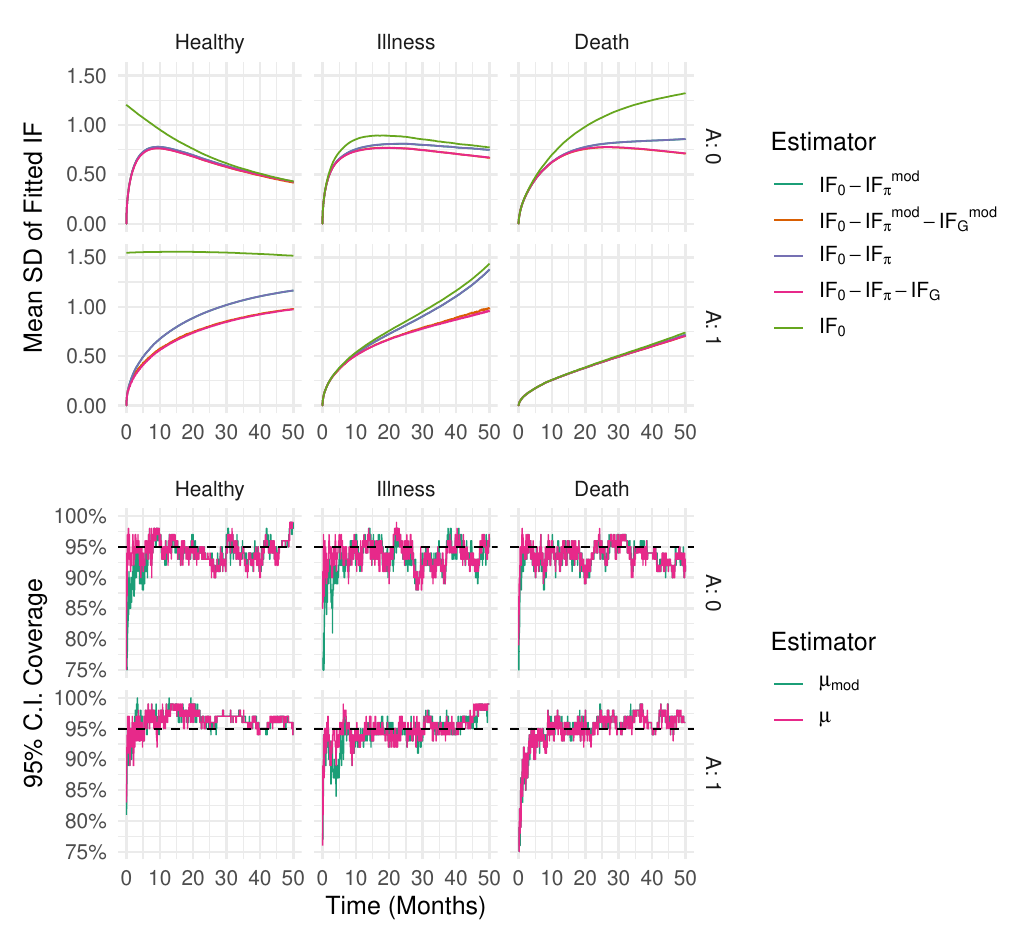}
        \caption{Mean SD for the influence functions of the estimators $\hat{\psi}_{n}^{0}$, $\hat{\psi}_{n}^{\nu}$, $\hat{\psi}_{n}^{\mu}$, $\hat{\psi}_{n}^{\nu, mod}$ and $\hat{\psi}_{n}^{\mu, mod}$ using correctly specified nuisance parameter estimators. Coverage of the corresponding 95\% confidence intervals for the estimators $\hat{\psi}_{n}^{\mu}$ and $\hat{\psi}_{n}^{\mu, mod}$. Averages are over $100$ independent datasets with a population size of $9000$ in each.}
        \label{fig:Test1_n9000_MSD_Cvg}
    \end{figure}
    
    In each dataset, we calculate the standard deviation (SD) of the fitted terms $IF_0$, $IF_0 - IF_{\pi}$ and $IF_0 - IF_{\pi} - IF_G$, to examine the hierarchy between them. The mean SD and corresponding 95\% coverage results for $\hat{\psi}_{n}^{\mu}$ and $\hat{\psi}_{n}^{\mu, mod}$ across the $100$ independent datasets are presented in Figure \ref{fig:Test1_n9000_MSD_Cvg}. The mean SD plots agree with the theoretical hierarchy outlined in Section \ref{subsec:prop}, and the resulting $95\%$ confidence intervals have approximately $95\%$ coverage, as expected.

\newpage
    
    As mentioned in \cite{Laan2003}, when the models for $\pi_{0}$ and $G_{0}$ are correctly specified, the regression coefficients from the modified estimators can be used as diagnostic tools for assessing the fit of the nuisance estimators $\hat Q_0$ and $\hat \eta_0$. We present statistics on the regression coefficients in Appendix \ref{app:mod_coef_diagnostics}. The results in Figure \ref{fig:Test1_ModCoefs} imply that the estimator $\hat Q_0$ is consistent, because $\beta^{*,\nu}_{\pi}(t) \approx  \beta^{*,\mu}_{\pi}(t) \approx 1 \; \forall t$. The coefficient $\beta^{*,\mu}_{G}(t)$ is also near $1$ for the \textit{Healthy} and \textit{Illness} states, but only converges to $1$ over time in the \textit{Death} state. This is expected, because the time-dependent variable dependence in $\hat \eta_0$ cannot be well estimated until several transitions have happened. We believe $\hat \eta_0$ could have a slightly better fit using interaction terms with $A$, but at a population size of $9000$ and many regression terms, we run into technical limitations. We have conducted a test in Figure \ref{fig:Test5_RMSE_ModCoefs} with interaction terms added to $\hat \eta_0$, run on $100$ simulations of $5000$ individuals, which shows a near perfect fit with of $\beta^{*,\mu}_{G}(t) \approx 1$ and slightly better RMSE gains than Figure \ref{fig:Test1_n9000_MBE_RMSE}. 
    
\subsection{Testing Robustness to Nuisance Estimation}
\label{subsec:misspecified}

To illustrate robustness properties, we use the same $100$ datasets, but with slightly biased estimators of the individual nuisance parameters. We misspecify each nuisance estimator by omitting variables, and choosing an incorrect functional form. The true models' formulas are functions of $f_1(Age)$ and $f_2(BMI)$, which are non-linear. In each case, we misspecify just one nuisance parameter at a time. 

\begin{figure}[H]
    \centering
    \includegraphics[width=\linewidth]{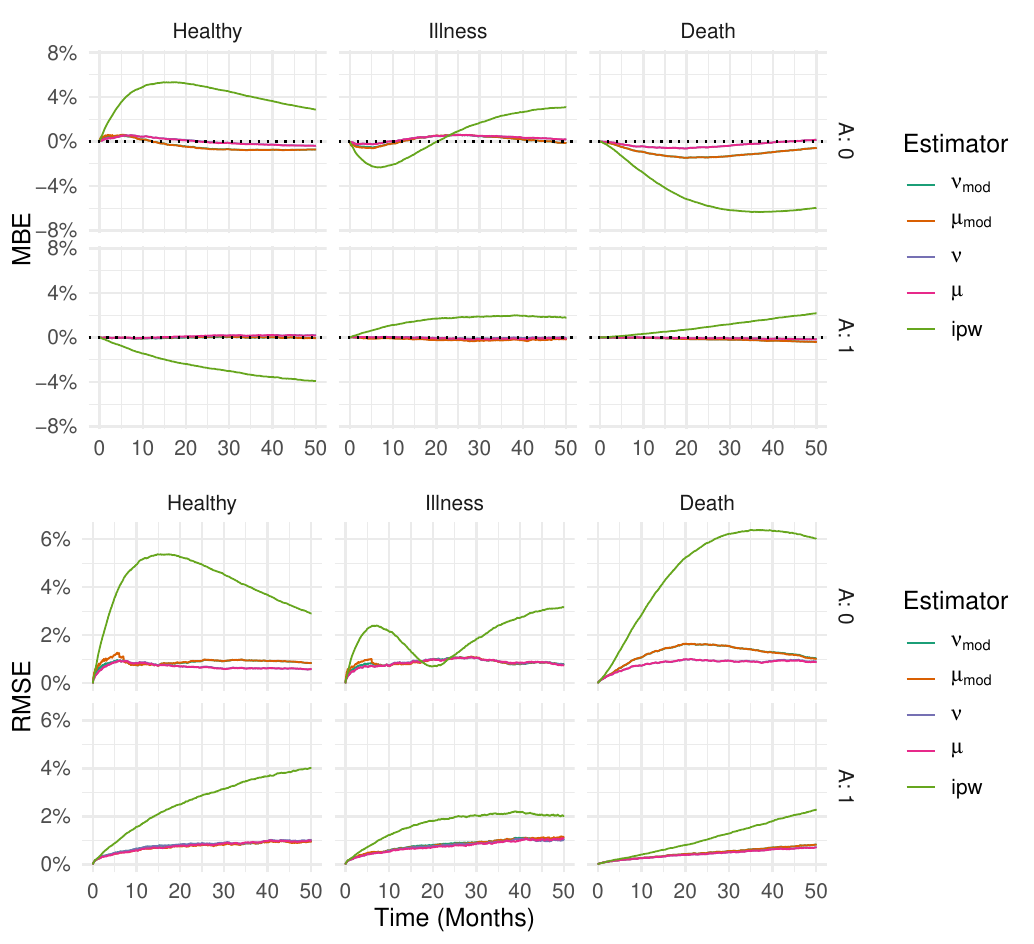}
    \caption{\textbf{Robustness to $\pi$}: MBE and RMSE metrics of the estimators $\hat{\psi}_{n}^{0}$, $\hat{\psi}_{n}^{\nu}$, $\hat{\psi}_{n}^{\mu}$, $\hat{\psi}_{n}^{\nu, mod}$ and $\hat{\psi}_{n}^{\mu, mod}$, all using the biased nuisance estimator $\hat \pi_{1}$. Averages are over $100$ independent datasets with a population size of $9000$ in each.}
    \label{fig:Test2a_MBE_RMSE}
\end{figure}

\begin{figure}[H]
    \centering
    \includegraphics[width=\linewidth]{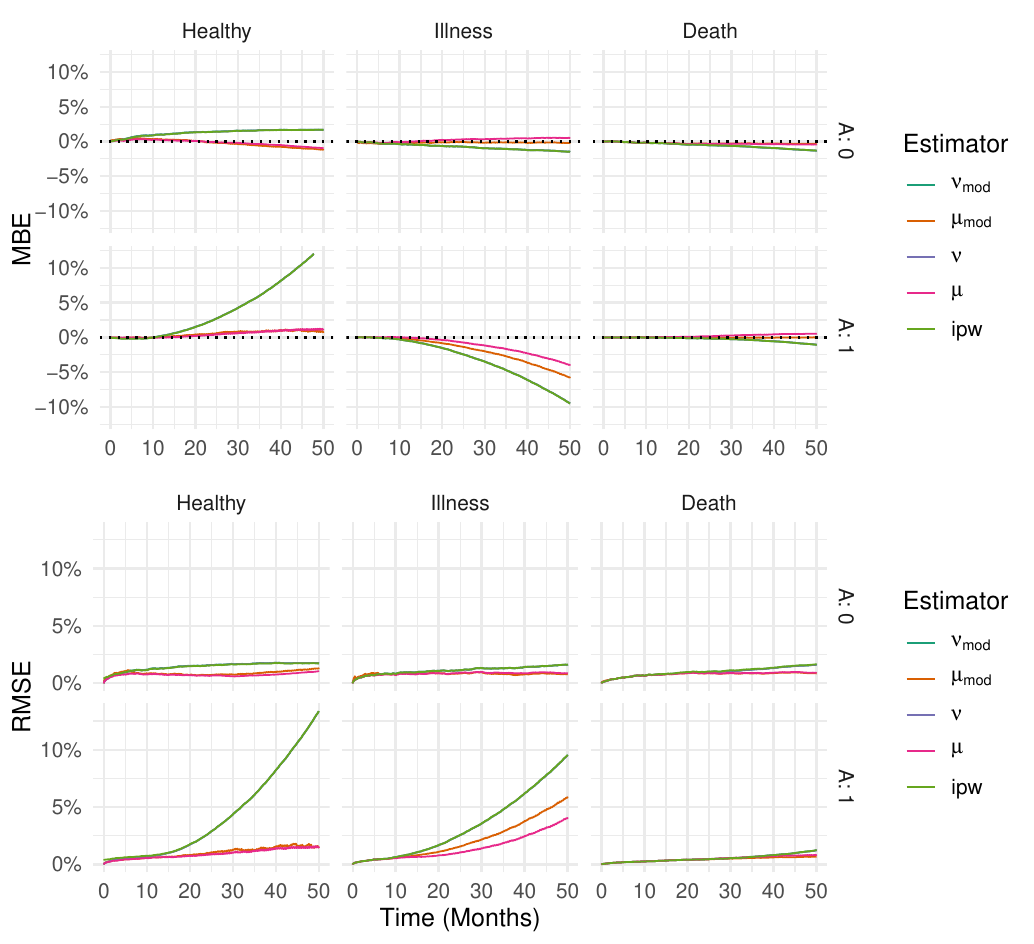}
    \caption{\textbf{Robustness to $G$}: MBE and RMSE metrics of the estimators $\hat{\psi}_{n}^{0}$, $\hat{\psi}_{n}^{\nu}$, $\hat{\psi}_{n}^{\mu}$, $\hat{\psi}_{n}^{\nu, mod}$ and $\hat{\psi}_{n}^{\mu, mod}$, all using the biased nuisance estimator $ \hat \lambda^G_{1a}$. Averages are over $100$ independent datasets with a population size of $9000$ in each.}
    \label{fig:Test3b_MBE_RMSE}
\end{figure}

\begin{figure}[H]
    \centering
    \includegraphics[width=\linewidth]{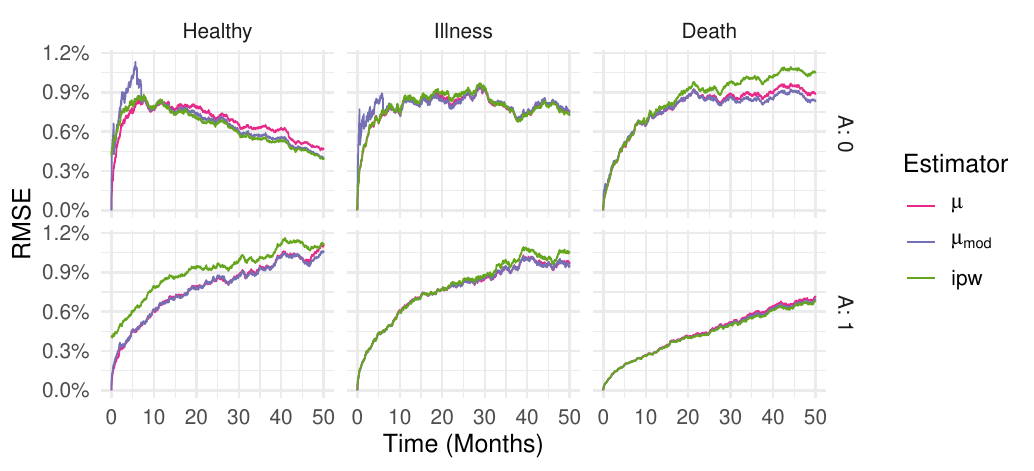}
    \caption{\textbf{Robustness to $\eta$}: RMSE for the estimators $\hat{\psi}_{n}^{0}$,  $\hat{\psi}_{n}^{\mu}$, and $\hat{\psi}_{n}^{\mu, mod}$, all using the biased nuisance estimator $\hat \eta_1$.  Averages are over $100$ independent datasets with a population size of $9000$ in each.}
    \label{fig:Test4b_RMSE}
\end{figure}

\begin{enumerate}
    \item \textbf{Robustness to $\pi_0$ estimation: }We define a biased estimator $\hat \pi_{1}$ for the true propensity score $\pi_0$ \eqref{eq:pi_correct_form} by using an incorrect model formula in the logistic regression. We omit all variables to ensure $\hat{\psi}_{n}^{0}$ is biased enough to analyse the robustness qualities of $\hat{\psi}_{n}^{\nu}$ and $\hat{\psi}_{n}^{\mu}$.
\begin{align}
 \hat \pi_{1}(1|w) &:= \text{expit}\left(\hat \xi_0  \right)
\end{align}
 The MBE and RMSE results for the estimators $\hat{\psi}_{n}^{0}$, $\hat{\psi}_{n}^{\nu}$, $\hat{\psi}_{n}^{\mu}$, $\hat{\psi}_{n}^{\nu, mod}$ and $\hat{\psi}_{n}^{\mu, mod}$ using $\hat \pi_{1}$ are presented here in Figure \ref{fig:Test2a_MBE_RMSE}. \newline

  \item \textbf{Robustness to $G_0$ estimation: }We define a biased estimator $\hat \lambda^G_{1a}(t)$ for the true censoring hazard $\lambda^{G_0}(t)$ \eqref{eq:G_correct_form} by omitting time-dependent variables and and using an incorrect linear form of \textit{Age} and \textit{BMI} in the Cox regression.
    \begin{align}
     \hat \lambda^G_{1a}(t) & := \hat \beta_0 \exp ( \hat \beta_1 A + \hat \beta_2 \textit{Age} + \hat \beta_3 \textit{BMI} )
    \end{align}
    
The MBE and RMSE results for the estimators $\hat{\psi}_{n}^{0}$, $\hat{\psi}_{n}^{\nu}$, $\hat{\psi}_{n}^{\mu}$, $\hat{\psi}_{n}^{\nu, mod}$ and $\hat{\psi}_{n}^{\mu, mod}$ using $\hat \lambda^G_{1a}$ are presented in Figure \ref{fig:Test3b_MBE_RMSE}. An additional test using an alternative misspecified estimator of $\lambda^{G_0}(t)$ is presented in Appendix \ref{app:subsec:moreplots}. \newline

    \item \textbf{Robustness to $Q_0$ and $\eta_0$ estimation: } We define biased nuisance estimators $\hat Q_{1}(t)$ and $\hat \eta_{1}(t,u)$ by omitting all variables except \textit{Sex} from the linear regressions:
    \begin{align}
     \hat \eta_1(t,u) := \hat \zeta_0(t,u) + \hat \zeta_1(t,u) Sex, \text{ and} && \hat Q_1(t) := \hat \eta_1(t,0) && \forall u \leq t.\label{eq:misspec_Q_eta}
    \end{align}
    
The RMSE results for the estimators $\hat{\psi}_{n}^{0}$, $\hat{\psi}_{n}^{\mu}$ and $\hat{\psi}_{n}^{\mu, mod}$ using $\hat \eta_{1}$ are presented in Figure \ref{fig:Test4b_RMSE}. The mean SD of the influence functions using either $\hat \eta_{1}$ or $\hat Q_{1}$ and RMSE in the latter case are presented in Figures \ref{fig:Test4a_MSD_RMSE}, and \ref{fig:Test4b_MSD} in Appendix \ref{subsec:sens_to_Q_eta}. MBE results are omitted, because they are all $\approx 0$.

\end{enumerate}

When using a biased nuisance estimator for $\pi_0$ or $G_0$, the estimator $\hat{\psi}_{n}^{0}$ returns a biased estimate for the occupation probability, at varying magnitudes. When $\lambda^{G_0}$ and $\pi_0$ are consistently estimated, using a biased nuisance estimator for $\eta_0$ or $Q_0$ should not cause bias in the occupation probability estimators, but they would not be efficient. The estimators $\hat{\psi}_{n}^{\nu, mod}$ and $\hat{\psi}_{n}^{\mu, mod}$ should account for this loss of efficiency, as explained in Section \ref{subsec:prop}.

\newpage 

The IPW estimator $\hat{\psi}_{n}^{0}$ is significantly biased, due to the biased nuisance estimator $\hat \pi_1$ and the IPW estimator's lack of robustness properties. The AIPW estimators, however, are all robust to the bias through their augmenting terms. Therefore, as expected, $\hat{\psi}_{n}^{\nu}$, $\hat{\psi}_{n}^{\mu}$, $\hat{\psi}_{n}^{\nu, mod}$ and $\hat{\psi}_{n}^{\mu, mod}$ all return MBE curves around $0$ for all six plots in Figure \ref{fig:Test2a_MBE_RMSE}. Their RMSE is also significantly less than the RMSE for the IPW estimator $\hat{\psi}_{n}^{0}$, as expected from the lack of bias. There is a short interval in the \textit{Illness} case without treatment, where RMSE for the $\hat{\psi}_{n}^{0}$ estimator is lower, because it briefly intersects $\psi_0(t)$, but then continues to trail off. Censoring is not misspecified, so $\hat{\psi}_{n}^{\mu} \approx \hat{\psi}_{n}^{\nu}$, because the $IF_G$ projection term is negligible compared to $IF_{\pi}$ in this test.

\newpage

\vspace{-4pt} 

The IPW estimator $\hat{\psi}_{n}^{0}$ is significantly biased, due to the biased nuisance estimator $\hat \lambda^G_{1a}$ and the IPW estimator's lack of robustness properties. The AIPW estimators $\hat{\psi}_{n}^{\nu}$ and $\hat{\psi}_{n}^{\nu, mod}$ perform similarly to the IPW estimator, as expected, because they are only partially augmented to adjust for bias in the nuisance parameter for treatment, not censoring. 

The two fully augmented estimators are expected to perform better, but we do not expect true robustness in practice. The estimator $\hat \eta_0$ uses a simplified linear form, and the observations in the regressions are weighted through biased weights from $\hat \lambda^G_{1a}$, which bias the fitted results. The magnitude of bias in the fitted censoring weight values $\hat{G}_{1a}(t)$ increases with time, but despite this, we see that $\hat{\psi}_{n}^{\mu, mod}$ and especially $\hat{\psi}_{n}^{\mu}$ are not significantly effected. Even in the treated \textit{Illness} case, where they begin to trail off, the unmodified fully augmented estimator $\hat{\psi}_{n}^{\mu}$ has less than half the bias of the IPW estimator $\hat{\psi}_{n}^{0}$. It is an open question how much remaining bias could be eliminated through a non-linear regression in the estimator $\hat \eta_0$ to better capture the true form of $\eta_0$. It may also be due to positivity violations in censoring weight values.

\newpage

We present the Figures for robustness to $\eta$ and $Q$ in Appendix \ref{subsec:sens_to_Q_eta}. When using the biased nuisance estimator $\hat \eta_1$, we see a slight improvement in the RMSE for $\hat{\psi}_{n}^{\mu, mod}$ compared to $\hat{\psi}_{n}^{\mu}$ in some plots, such as the \textit{Healthy} and \textit{Death} state estimates in the top row of Figure \ref{fig:Test4b_RMSE} above. We see a similar RMSE improvement for two plots in Figure \ref{fig:Test4a_MSD_RMSE} when using the estimator $\hat Q_1$, but no impact on the other four. We see a more consistent impact of the modified estimators on the mean SD of the fitted influence functions when using the biased $\hat Q_1$ estimator, in Figure \ref{fig:Test4a_MSD_RMSE}. It is possible that $\hat{\psi}_{n}^{\mu, mod}$ would have a more significant impact in other frameworks, or with different biased estimators of $Q_0$ and $\eta_0$. Regardless, as a diagnostic test, the results are aligned with what we expect from the theory. 

Additionally, we can assess the modifying regression coefficients in Appendix \ref{app:mod_coef_diagnostics}. In the correctly specified case with nuisance estimators $\hat Q_0$ and $\hat \eta_0$, the regression coefficients were approximately 1. In contrast, when using the estimators $\hat Q_1$ or $\hat \eta_1$ we see in Figures \ref{fig:Test4a_ModCoefs} and \ref{fig:Test4b_ModCoefs} that the coefficients are much further off. With the $\hat \eta_1$ estimator, in the untreated \textit{Death} state, $\beta^{*,\mu}_{G}$ does not even converge into the interval [-2,2], which implies that it gives a heavily biased estimate of $\eta_0$. We can draw a similar conclusion about the consistency of the $\hat Q_1$ estimator from the $\beta^{*,\nu}_{\pi}$ and $\beta^{*,\mu}_{\pi}$  coefficients.

\newpage

\section{Discussion} 
\label{sec:discussion}

\subsection{Estimating hazards or occupation probabilities}
\label{subsec:disc:haz_occprob}
In case specific multistate models robust methods have been proposed, see e.g. \cite{Martinussen2023}, \cite{Rytgaard2023}. Different from our approach, these authors consider the parameter of interest as a functionals of transition hazards functions. In the general multi-state setting the number of transitions can vastly outnumber the number of states. A saturated model with $K$ states will have $K^{2} - K$ transitions. In so far as we are concerned about model misspecification, estimating transition probabilities using our proposed AIPW estimators may be a desirable alternative, resulting in (in addition to propensity scores and censoring kernel) $2K$ parameters.

\subsection{Discussion of censoring assumption}
\label{subsec:disc:cens}
An important feature of the weaker coarsening assumptions is that right censoring is allowed to depend on the prior history of the multi-state process (and other covariates process). While coarsening assumptions should be justified on a case by case basis, it seems only reasonable that if a larger multi-state process is necessary to represent a given real world phenomena, such as an illness-death model with recurrence, then this is already an admission that path-dependencies should be thought about carefully and right censoring should be allowed to depend on at least the internal history of the process. A common censoring assumption in the simple survival and competing risk models, is independence of the latent outcomes given baseline covariants. This is a technically convenient assumption in that the representation of $\eta$ simplifies significantly and the dependence of $G$ can be avoided (see e.g., \cite{Scheike2020}, \cite{CaiLaan2020}). However, including time dependent covariates in the coarsening assumptions introduces these problems again as also discussed in \cite{Hubbard2000}. The resulting positivity assumption under CAR on the censoring kernel implied by \eqref{eq:ObsPos} is a strict assumption, and for it to hold for a censoring model in a finite sample, some dimensional reduction is necessary. This is a common problem for identification in longitudinal settings (see e.g., \cite{KKN2021}) but it is particularly relevant to larger multi-models since by design the histories are allowed to be huge. While the coarsening assumptions in Section \ref{sec:EffInfl} are weaker than assumptions commonly seen in the literature on causal inference in continuous time to event settings, one could (and perhaps should) question the CAR assumptions as realistic data generating mechanism.,  compared. As pointed out by \citep{baer2023causal} CAR imply so-called cross world assumptions unlike standard sequential exchangeability assumptions in causal inference. While acknowledging these shortcomings, we should remember that a central reason introducing the coarsening at random assumption historically was as a tool for ensuring properties allowing likelihood based inference as discussed in \cite{RubinHeitjan1991} and \cite{Heitjan1997} and for constructing semi-parametric models with maximal score spaces in general missing data models as discussed in \cite{GLR1997}.

\subsection{Simulation experiment and implementation} 
\label{subsec:disc:sim}
The findings of the simulation experiments are in agreement with the theoretical properties of the estimators. Using consistent estimators of the nuisance parameters, the $\mu$ estimator converged more efficiently than the IPW estimator. When the estimators for the propensity score or censoring weights were biased, the IPW estimator was heavily biased, while the $\mu$ estimator was robust in both cases. Thus, we recommend usage of the $\mu$ estimator for datasets with sufficient data and significant censoring. We presented performance results on datasets with $9000$ individuals, but running the same tests on smaller datasets of $2500$ individuals yields similar results. The main difference is in the estimator $\hat{\psi}_{n}^{\mu, mod}$, which shows more instability, especially at early times. The $\mu$ estimator does not break down unless the observation count is much lower. We must emphasize, however, that the IPW estimator and $\nu$ estimators are clearly more stable. To obtain the $\mu(t)$ estimate for time $t$, we must estimate the $\eta(t,u)$ values at every censoring time $u \leq t$, and integrate them over $u$. This requires sufficient individuals to be at risk in every strata at all times $u$ to calibrate coefficients for all variables and interaction terms. If this is not satisfied, it is still worth using $\nu$ over the IPW estimator, due to superior efficiency and robustness properties. While the coefficients estimates of the modified estimators serve as a diagnostic tools, useful in particular to asses the fit of $\eta$, more flexible and adaptive methods would be preferable if feasible under the constraints \eqref{MG_assumption}. 

The results for the final plots in Section \ref{sec:4:simulations} were computed using the med-biostat2 HPC server, provided by the Oslo Centre for Epidemiology and Biostatistics, at the University of Oslo. However, the R code is parallelised and optimised to run on weaker personal computers at smaller, but significant sample sizes. For a dataset of $2500$ individuals, the $\mu$ estimator takes only around $20$ seconds to run on a MacBook Pro (14-inch, 2021) with 32GB memory and an Apple M1 Pro chip with 8 CPU cores.

\subsection*{Conflict of interest}
There are no financial or commercial conflict of interest to report.

\subsection*{Data availability}
All R-code used to construct the simulation experiment is available on \texttt{https://github.com/lukatsgd/multistate-AIPW}. We aim to publish the estimator codes as a package in the future.

\subsection*{Acknowledgements}
This research was funded by The Norwegian Research Council, project number 315323.

\bibliographystyle{biometrika}
\bibliography{paper-ref}

\newpage

\appendix
\appendixone

\section{(Coarsening at random)}
\label{app:A:CAR}

Sufficient conditions for CAR under baseline exposure and right censoring are given in \cite{Hubbard2000}, and it suffices to verify that these are satisfied. Assumptions \eqref{eq:CAR_A} and \eqref{eq:CAR_C1} implies 
\begin{align*}
    \prp(A=a \vert \bar{X}_{0}, \bar{X}_{1}) = \prp(A=a \vert W)
\end{align*}
which is their first condition. The condition \eqref{eq:CAR_C2} is equivalent to their second condition, that the hazard of $C_{a}$ given $\bar{X}_{a}$ at time $t$ is a function of $(t, \bar{X}_{a}(t))$, (see e.g., \cite{LR2003}[Section 3]).

\subsection*{Factorization of coarsening kernel:}
Let us realize the factorization of the coarsening kernel. First of all
\begin{align}
  A \ci_{\prp} \bar{X}_{a}, C_{a} \vert W \label{CIprim}
\end{align}
implies 
\begin{align*}
       \prp(A=a \vert \bar{X}_{a}, C_{a},  W) =  \prp(A=a \vert W) = \pi(a \vert W;\prp).
\end{align*}
Since $W$ is part of $\bar{X}_{a}$, weak union implies (see \cite{JKSKW2005} Corollary A.2, p.236) 
\begin{align}
  A \ci_{\prp} C_{a} \vert \bar{X}_{a}. \label{CIcens}
\end{align}
Therefore, 
\begin{align}
    \prp(A=a, C_{A} \in \diff t \vert \bar{X}_{a}) &= \prp(A=a  \vert \bar{X}_{a}) \prp(C_{a} \in \diff t \vert \bar{X}_{a}) =  \pi(a \vert W;\prp) G( \diff t \vert \bar{X}_{a}, a; \prp). \label{factorizeA1}
\end{align}
In order to see that $G( \diff t \vert x, a; \prp)$ determines the observable censoring events, as expressed by the equality \eqref{eq:ObsCFcens} in the main text i.e.  
\begin{align*}
   \prp(C_{a} \in \diff t \vert \bar{X}_{a}) &= \prp(C_{A} \in \diff t \vert \bar{X}_{a}, A= a) 
\end{align*}
we first define the meaning of the right hand side conditioning on a sigma algebra and a set. Here we follow \cite{Vaart2004} (see lemma A.8) and use the definition 
\begin{align*}
    \prp(C_{A} \in \diff t \vert \bar{X}_{a}, A= a) =  \frac{\prp(C_{a}  \in \diff t, A=a \vert \bar{X}_{a})}{\prp(A=a \vert \bar{X}_{a})}.
\end{align*}
By the factorization \eqref{factorizeA1} the right hand side above is $G( \diff t \vert \bar{X}_{a}, a; \prp)$. Furthermore, 
\begin{align*}
    \prp(C \in \diff t \vert \bar{X},A) &=  \sum_{a}\mathds{1}\{A=a\} \frac{\prp(C_{a}  \in \diff t, A=a \vert \bar{X}_{a})}{\prp(A=a \vert \bar{X}_{a})} \\
    &= \sum_{a} \mathds{1}\{A=a\} G( \diff t \vert \bar{X}_{a},a; \prp) = G( \diff t \vert \bar{X},A; \prp).
\end{align*}

\newpage

\appendixtwo

\section{Observed data influence function}
\label{app:B:inf_fn}

Define for fixed but arbitrary $a \in \{0,1\}$, which we will suppress in notation,
\begin{align*}
    B(C,\bar{X}) &:= \frac{\mathds{1}\{A=a\}\Delta^{t_{0}} H}{G_{0}(\tilde{\tau} \wedge t_{0} \vert \bar{X}, a)\pi_{0}(a \vert W)} - \psi_{0}
\end{align*}
As explained in the main text we identify the observed data efficient influence function using the projection formula (see e.g. theorem 4.4 of \cite{GLR1999} or Theorem 1.3 of \cite{LR2003}). More specifically, we will identify it as $B(C,\bar{X}) - \Pi(B(C,\bar{X}) \vert \mathbb{T}_{CAR})$, where $ \Pi(B(C,\bar{X}) \vert \mathbb{T}_{CAR})$ is the projection of $B(C,\bar{X})$ onto $\mathbb{T}_{CAR}$. By the factorization $\mathbb{T}_{CAR} = \mathbb{T}_{\pi} \oplus \mathbb{T}_{G}$, we have that $\Pi(B(C,\bar{X}) \vert \mathbb{T}_{CAR}) = \Pi(B(C,\bar{X}) \vert \mathbb{T}_{G}) + \Pi(B(C,\bar{X}) \vert \mathbb{T}_{\pi})$ and the goal is therefore to identify each of these projections. To utilize this strategy, we first verify that $\e_{0}[ B(C,\bar{X}) \vert \bar{X}_{a}]$ identifies the influence function of the full data parameter $\psi_{\bar{\bm{X}}}(F) := \e_{F,G,\pi}[H_{a}]$ where $H_{a}:=\mathds{1}\{Z_{a}(\tau_{a} \wedge t_{0}) = j\}$. To that end, observe that on the event $\{\Delta^{t_{0}} = 1\}$ we have $\tilde{\tau} = \tau$ and going forward we will replace $\tilde{\tau}$ by $\tau$ when taking expectations of $B(C,\bar{X})$. First, observe that 
\begin{align*}
    \e_{0}[\mathds{1}\{A=a\}\Delta^{t_{0}} \vert \bar{X}_{a}] &= \e_{0}[\mathds{1}\{A=a\}\Delta_{a}^{t_{0}} \vert \bar{X}_{a}] \\
    &= \prp_{0}(A=a, C_{a}> \tau_{a} \wedge t_{0} \vert \bar{X}_{a}) \\
    &= \prp_{0}(C_{a}> \tau_{a} \wedge t_{0} \vert \bar{X}_{a}) \prp_{0}(A=a \vert \bar{X}_{a}) \\ 
    &= G_{0}(\tau_{a} \wedge t_{0} \vert \bar{X}_{a}, a)\pi_{0}(a \vert W)
\end{align*}
Therefore,
\begin{align*}
    \e_{0}[ B(C,\bar{X}) \vert \bar{X}_{a}] &= \e_{0}\left[ \frac{\mathds{1}\{A=a\}\Delta_{a}^{t_{0}} H_{a}}{G_{0}(\tau_{a} \wedge t_{0} \vert \bar{X}_{a}, a)\pi_{0}(a \vert W)} \Big\vert \bar{X}_{a} \right] - \psi_{0} \\
    &= \frac{H_{a} \e_{0}[\mathds{1}\{A=a\}\Delta_{a}^{t_{0}} \vert \bar{X}_{a}]}{G_{0}(\tau_{a} \wedge t_{0} \vert \bar{X}_{a}, a)\pi_{0}(a \vert W)}   - \psi_{0} = H_{a} - \psi_{0}
\end{align*}
which we recognize as the full data efficient influence function of $\psi_{\bar{\bm{X}}}(F)$ at $F_{0}$.

\subsection{Versions of $Q$ and $\eta$}

In the main text, we used particular representations of $Q$ and $\eta$ based on observed data. There are multiple equivalent representations of these nuisance parameters, and for later calculations the following representations will be convenient:  
\begin{align}
    Q_{0}(t_{0}, a,W) &= \e_{0}[H \vert A=a, W] \label{Qdef} \\
    \eta_{0}(t_{0}, u, \bar{X}(u), a) &= \e_{0}[H \vert \bar{X}(u), \tilde{\tau} > u, A=a] \label{etadef}.
\end{align}
The equivalence between the above representations and those in the main text will be justified later. 

\subsection{Projection to $\mathbb{T}_{\pi}$}

We use that the projection $\Pi(B(C,\bar{X}) \vert \mathbb{T}_{\pi})$ is given by
\begin{align}
    \Pi(B(C,\bar{X}) \vert \mathbb{T}_{\pi}) = \e_{0}[B(C,\bar{X}) \vert A, W] - \e_{0}[B(C,\bar{X}) \vert W]. \label{piProjRep}
\end{align}
For a justification see e.g. \cite{LR2003}. To identify $\e_{0}[B(C,\bar{X}) \vert A, W]$ observe that 
\begin{align*}
     \e_{0} \left[\frac{\mathds{1}\{A=a\}\Delta_{a}^{t_{0}} H_{a}}{G_{0}(\tau_{a} \wedge t_{0} \vert \bar{X}_{a}, a)\pi_{0}(a \vert W) }  \Big\vert A, W \right] &= \e_{0}\left[\e_{0} \left[\frac{\mathds{1}\{A=a\}\Delta_{a}^{t_{0}} H_{a}}{G_{0}(\tau_{a} \wedge t_{0} \vert \bar{X}_{a}, a)\pi_{0}(a \vert W) }  \Big\vert \bar{X}_{a}, A\right] \vert A, W \right] \\
     &= \frac{\mathds{1}\{A=a\} }{ \pi_{0}(a \vert W) } \e_{0}\left[ \frac{ H_{a}}{ G_{0}(\tau_{a} \wedge t_{0} \vert \bar{X}_{a}, a) } \e_{0} \left[ \Delta_{a}^{t_{0}} \Big\vert \bar{X}_{a}, A\right] \vert A, W \right]. 
\end{align*}
By \eqref{CIcens} have $\e_{0} \left[ \Delta_{a}^{t_{0}} \Big\vert \bar{X}_{a}, A\right] = \e_{0} \left[ \Delta_{a}^{t_{0}} \Big\vert \bar{X}_{a} \right] = G_{0}(\tau_{a} \wedge t_{0} \vert \bar{X}_{a}, a)$ and by \eqref{CIprim} we have 
\begin{align*}
\e_{0} \left[ H_{a} \Big\vert W, A\right] &= \e_{0} \left[ H_{a} \Big\vert W\right] 
\end{align*}
Again by \eqref{CIprim} we have 
\begin{align*}
\e_{0} \left[ H_{a} \Big\vert W \right] &= \e_{0} \left[ H_{a} \Big\vert W, A=a \right] \\
&= \frac{\e_{0}[\mathds{1}\{A=a\}H_{a} \vert W]}{\e_{0}[\mathds{1}\{A=a\} \vert W]} \\
&= \frac{\e_{0}[\mathds{1}\{A=a\}H \vert W]}{\e_{0}[\mathds{1}\{A=a\} \vert W]}= \e_{0} \left[ H \Big\vert W, A=a \right] = Q_{0}(a, W)    
\end{align*}
We conclude that 
\begin{align*}
    \e_{0}[B(C,\bar{X}) \vert A, W] = \frac{\mathds{1}\{A=a\} }{ \pi_{0}(a \vert W) } Q_{0}(a,W)  .
\end{align*}
Using iterated expectations we get that 
\begin{align*}
    \e_{0}[B(C,\bar{X}) \vert A, W] - \e_{0}[B(C,\bar{X}) \vert W] = \left(\frac{\mathds{1}\{A=a\} }{ \pi_{0}(a \vert W) } -1\right)Q_{0}(a,W).
\end{align*}
We proceed by verifying that the version of $Q_{0}$ in equation \eqref{eq:Q} of the main text is equivalent to \eqref{Qdef}. To do so, we use from the calculations above that $\e_{0} \left[ H_{a} \Big\vert W \right] = Q_{0}(a,W)$. Furthermore, we have 
\begin{align*}
    \e_{0} \left[ H_{a} \Big\vert W \right] &= \e_{0} \left[ \e_{0} \left[ \frac{\Delta_{a}^{t_{0}} H_{a}}{G_{0}(\tau_{a} \wedge t_{0} \vert \bar{X}_{a}, a)}  \Big\vert \bar{X}_{a} \right]   \Big\vert W \right] = \e_{0} \left[ \frac{\Delta_{a}^{t_{0}} H_{a}}{G_{0}(\tau_{a} \wedge t_{0} \vert \bar{X}_{a}, a)}  \Big\vert W \right]
\end{align*}
and by \eqref{CIprim} we have 
\begin{align*}
    \e_{0} \left[ \frac{\Delta_{a}^{t_{0}} H_{a}}{G_{0}(\tau_{a} \wedge t_{0} \vert \bar{X}_{a}, a)}  \Big\vert W \right] &= \e_{0} \left[ \frac{\Delta_{a}^{t_{0}} H_{a}}{G_{0}(\tau_{a} \wedge t_{0} \vert \bar{X}_{a}, a)}  \Big\vert A=a, W \right] \\
    &= \frac{\e_{0} \left[\mathds{1}\{A=a\} \frac{\Delta_{a}^{t_{0}} H_{a}}{G_{0}(\tau_{a} \wedge t_{0} \vert \bar{X}_{a}, a)} \Big\vert W \right]}{\e_{0}[\mathds{1}\{A=a\} \vert W]} \\
    &= \frac{\e_{0} \left[\mathds{1}\{A=a\} \frac{\Delta^{t_{0}} H }{G_{0}(\tau \wedge t_{0} \vert \bar{X}, a)} \Big\vert W \right]}{\e_{0}[\mathds{1}\{A=a\} \vert W]} \\
    &= \e_{0} \left[ \frac{\Delta^{t_{0}} H }{G_{0}(\tau \wedge t_{0} \vert \bar{X}, a)}  \Big\vert A=a, W \right].
\end{align*}
 
\subsection{Projection to $\mathbb{T}_{G}$}

In order to show that martingale term $IF_{G}$ in equation \eqref{MGtermIF} of the main text is the projection of $B(C,\bar{X})$ onto $\mathbb{T}_{G}$ we use \cite{GLR1999} Theorem 5.2 (see also \cite{Hubbard2000} Proposition 5.1 and \cite{Vaart2004} Theorem 2.2) which states that the projection is the martingale integral with integrand
\begin{align*}
    &\e_{0}\left[ \frac{\mathds{1}\{A=a\}}{\pi_{0}(a \vert W)}\frac{H\{\Delta_{\tau,u}^{t_{0}} -\Delta_{\tau,C}^{t_{0}}\}}{G_{0}(\tau \wedge t_{0} \vert \bar{X}, a)} \Big \vert \bar{X}(u), A, C > u \right], 
\end{align*}
where $\Delta_{\tau,C}^{t_{0}} = \mathds{1}\{C > \tau \wedge t_{0}\}$ and $\Delta_{\tau,u}^{t_{0}} = \mathds{1}\{u > \tau \wedge t_{0}\}$. According to the representation \eqref{etadef} of $\eta_{0}$, it suffices to show that for $u \leq t_{0}$
\begin{align}
    &\e_{0}\left[ \frac{\mathds{1}\{A=a\}}{\pi_{0}(a \vert W)}\frac{H\{\Delta_{\tau,u}^{t_{0}} -\Delta_{\tau,C}^{t_{0}}\}}{G_{0}(T \vert \bar{X},a)} \Big \vert \bar{X}(u), A, C > u \right] = \label{main_eq} \\
    &- \frac{\mathds{1}\{\tau \geq u, A=a\}}{\pi_{0}(a, W)G_{0}(u \vert \bar{X}, a)} \e_{0}\left[ H \vert \bar{X}(u), \tilde{\tau} \geq u, A= a \right]. \nonumber
\end{align}
We will subsequently show how to rewrite $\eta$ according to the definition in the main text. Focusing on \eqref{main_eq} we have  
    \begin{align}
        &\e_{0}\left[ \frac{\mathds{1}\{A=a\}}{\pi_{0}(a \vert W)}\frac{H\{\Delta_{\tau,u}^{t_{0}} -\Delta_{\tau,C}^{t_{0}}\}}{G_{0}(\tau \wedge t_{0} \vert \bar{X},a)} \Big \vert \bar{X}(u),A, C > u \right] = \nonumber \\
        &\frac{\mathds{1}\{A=a\}}{\pi_{0}(a \vert W)} \e_{0}\left[ \frac{H_{a}\{\Delta_{\tau_{a},u}^{t_{0}} -\Delta_{\tau_{a},C_{a}}^{t_{0}}\}}{G_{0}(\tau_{a} \wedge t_{0} \vert \bar{X}_{a},a)} \Big \vert \bar{X}(u),A, C > u \right] = \nonumber \\
        &\frac{\mathds{1}\{A=a\}}{\pi_{0}(a \vert W)} \e_{0}\left[ \frac{H_{a}\{\Delta_{\tau_{a},u}^{t_{0}} -\Delta_{\tau_{a},C_{a}}^{t_{0}}\}}{G_{0}(\tau_{a} \wedge t_{0} \vert \bar{X}_{a},a)} \Big \vert \bar{X}_{a}(u),A = a, C_{a} > u \right] = \nonumber \\
        &\frac{\mathds{1}\{A=a\}}{\pi_{0}(a \vert W)} \e_{0}\left[ \frac{H_{a}\{\Delta_{\tau_{a},u}^{t_{0}} -\Delta_{\tau_{a},C_{a}}^{t_{0}}\}}{G_{0}(\tau_{a} \wedge t_{0} \vert \bar{X}_{a},a)} \Big \vert \bar{X}_{a}(u), C_{a} > u \right]  \label{expression1}
    \end{align}
    
    It remains to identify 
    \begin{align*}
        \e_{0}\left[ \frac{H_{a}\{\Delta_{\tau_{a},u}^{t_{0}} -\Delta_{\tau_{a},C_{a}}^{t_{0}}\}}{G_{0}(\tau_{a} \wedge t_{0} \vert \bar{X}_{a},a)} \Big \vert \bar{X}_{a}(u), C_{a} > u \right].
    \end{align*}
    This is done in exactly the same way as in \cite{GLR1999}. We start by using a generalised law of iterated expectations (see lemma \ref{lemma:iterexp_set} below), conditioning on $\bar{X}_{a}, C_{a} > u$ in the innermost conditional expectation. This innermost expectation is
    \begin{align*}
        &\e_{0}\left[ \frac{H_{a}\{\Delta_{\tau_{a},u}^{t_{0}} -\Delta_{\tau_{a},C_{a}}^{t_{0}}\}}{G_{0}(\tau_{a} \wedge t_{0} \vert \bar{X}_{a},a)} \Big \vert \bar{X}_{a}, C_{a} > u \right] = \\
         &\frac{H_{a}\left\{\Delta_{\tau_{a},u}^{t_{0}} - \e_{0}\left[ \Delta_{\tau_{a},C_{a}}^{t_{0}}  \vert \bar{X}_{a}, C_{a} > u \right] \right\}}{G_{0}(\tau_{a} \wedge t_{0} \vert \bar{X}_{a},a)} 
    \end{align*}
    Using the following representation 
    \begin{align}
        \e_{0}\left[\Delta_{\tau_{a},C_{a}}^{t_{0}}  \vert \bar{X}_{a}, C_{a} > u \right] &= \frac{\e_{0}\left[\mathds{1}\{ C_{a} > u \}  \mathds{1}\{ C_{a} > \tau_{a} \wedge t_{0} \}  \vert \bar{X}_{a} \right]}{\e_{0}\left[ \mathds{1}\{ C_{a} > u \}  \vert \bar{X}_{a} \right]} \nonumber \\
        &= \frac{\e_{0}\left[\mathds{1}\{ C_{a} > u \vee (\tau_{a} \wedge t_{0}) \}  \vert \bar{X}_{a} \right]}{\e_{0}\left[ \mathds{1}\{ C_{a} > u \}  \vert \bar{X}_{a} \right]} \nonumber \\
        &= \frac{G_{0}(u \vee (\tau_{a} \wedge t_{0}) \vert \bar{X}_{a},a)}{G_{0}(u \vert \bar{X}_{a},a)}, \label{WeightEq}
    \end{align}
    we get
    \begin{align*}
        &\e_{0}\left[ \frac{H_{a}\{\Delta_{\tau_{a},u}^{t_{0}} -\Delta_{\tau_{a},C_{a}}^{t_{0}}\}}{G_{0}(\tau_{a} \wedge t_{0} \vert \bar{X}_{a},a)} \Big \vert \bar{X}_{a}(u), C_{a} > u \right] = \\
        &\e_{0}\left[ \frac{H_{a}\left\{\Delta_{\tau_{a},u}^{t_{0}} - \frac{G_{0}(u \vee (\tau_{a} \wedge t_{0}) \vert \bar{X}_{a},a)}{G_{0}(u \vert \bar{X}_{a},a)} \right\}}{G_{0}(\tau_{a} \wedge t_{0} \vert \bar{X}_{a},a)} \Big \vert \bar{X}_{a}(u), C_{a} > u \right].     
    \end{align*}
    Now use 
    \begin{align*}
        \frac{G_{0}(u \vee (\tau_{a} \wedge t_{0}) \vert \bar{X}_{a},a)}{G_{0}(u \vert \bar{X}_{a},a)} &= \mathds{1}\{\tau_{a} \wedge t_{0} < u\} + \mathds{1}\{\tau_{a} \wedge t_{0} \geq u\}\frac{G_{0}(\tau_{a} \wedge t_{0}  \vert \bar{X}_{a},a)}{G_{0}(u \vert \bar{X}_{a},a)} 
    \end{align*}
to obtain
\begin{align}
        &\e_{0}\left[ \frac{H_{a}\{\Delta_{\tau_{a},u}^{t_{0}} -\Delta_{\tau_{a},C_{a}}^{t_{0}}\}}{G_{0}(\tau_{a} \wedge t_{0} \vert \bar{X}_{a},a)} \Big \vert \bar{X}_{a}(u), C_{a} > u \right] = \nonumber \\
        &- \e_{0}\left[ \frac{H_a}{G_{0}(\tau_{a} \wedge t_{0} \vert \bar{X}_{a},a)} \mathds{1}\{\tau_{a} \wedge t_{0} \geq u\}\frac{G_{0}(\tau_{a} \wedge t_{0} \vert \bar{X}_{a},a)}{G_{0}(u \vert \bar{X}_{a},a)} \vert \bar{X}_{a}(u), C_{a} > u \right] = \nonumber \\
    &- \frac{\mathds{1}\{\tau_{a} \wedge t_{0} \geq u\}}{G_{0}(u \vert \bar{X}_{a}, a)} \e_{0}\left[ H_{a}  \vert \bar{X}_{a}(u), C_{a} > u\right].  \label{mainEtaId}  
\end{align}
By $\sigma(\bar{X}_{a}(u))$-measurability of $\mathds{1}\{\tau_{a} \wedge t_{0} \geq u\}$ we have     
\begin{align*}
    &\mathds{1}\{\tau_{a} \wedge t_{0} \geq u\}\e_{0}\left[ H_{a}  \vert \bar{X}_{a}(u), C_{a} > u\right] = \\
    &\mathds{1}\{\tau_{a} \wedge t_{0} \geq u\} \frac{\e_{0}\left[ \mathds{1}\{\tau_{a} \wedge t_{0} \geq u, C_{a} > u\}  H_{a}  \vert \bar{X}_{a}(u)\right]}{\e_{0}\left[ \mathds{1}\{\tau_{a} \wedge t_{0} \geq u, C_{a} > u\}  \vert \bar{X}_{a}(u) \right]} = \\
    &\mathds{1}\{\tau_{a} \wedge t_{0} \geq u\} \e_{0}\left[ H_{a}  \vert \bar{X}_{a}(u), C_{a} > u, \tau_{a} \wedge t_{0} \geq u\right] = \\
    &\mathds{1}\{\tau_{a} \wedge t_{0} \geq u\} \e_{0}\left[ H_{a}  \vert \bar{X}_{a}(u), \tilde{\tau}_{a} \wedge t_{0} \geq u\right]
\end{align*}
where the last equality follows since $C_{a}$ is totally inaccessible and therefore has point mass zero at $u$. Furthermore, for $u \leq t_{0}$ then
\begin{align*}
    \e_{0}\left[ H_{a}  \vert \bar{X}_{a}(u), 
    \tilde{\tau}_{a} \wedge t_{0} \geq u\right] = \e_{0}\left[ H_{a}  \vert \bar{X}_{a}(u), \tilde{\tau}_{a} \geq u\right].
\end{align*}
We now have, for $u \leq t_{0}$, 
\begin{align*}
        &\e_{0}\left[ \frac{H_{a}\{\Delta_{\tau_{a},u}^{t_{0}} -\Delta_{\tau_{a},C_{a}}^{t_{0}}\}}{G_{0}(\tau_{a} \wedge t_{0} \vert \bar{X}_{a},a)} \Big \vert \bar{X}_{a}(u), C_{a} > u \right] = \\
        &- \frac{\mathds{1}\{\tau_{a}  \geq u\}}{G_{0}(u \vert \bar{X}_{a}, a)}\e_{0}\left[ H_{a}  \vert \bar{X}_{a}(u), \tilde{\tau}_{a} \geq u\right]   
\end{align*}
The last step is to identify the conditional expectation above from observable data. Recall from \eqref{expression1} that we are multiplying by $\mathds{1}\{A=a\}$ therefore, we may replace $\mathds{1}\{\tau_{a} \geq u\} / G_{0}(u \vert \bar{X}_{a}, a)$ by $\mathds{1}\{\tau  \geq u\} / G_{0}(u \vert \bar{X}, a)$. Using conditional independence \eqref{CIprim} we have 
\begin{align*}
    \e_{0}\left[ H_{a}  \vert \bar{X}_{a}(u), \tilde{\tau}_{a} \geq u\right] &= \e_{0}\left[ H_{a}  \vert \bar{X}_{a}(u), \tilde{\tau}_{a} \geq u, A= a\right] \\
    &= \frac{\e_{0}\left[ \mathds{1}\{\tilde{\tau} \geq u, A= a\} H  \vert \bar{X}_{a}(u) \right]}{\e_{0}\left[ \mathds{1}\{\tilde{\tau} \geq u, A= a\}  \vert \bar{X}_{a}(u) \right]} \\
\end{align*}
Use that we are multiplying by $\mathds{1}\{A=a\}$ and on that event we have 
\begin{align*}
    \frac{\e_{0}\left[ \mathds{1}\{\tilde{\tau} \geq u, A= a\} H  \vert \bar{X}_{a}(u) \right]}{\e_{0}\left[ \mathds{1}\{\tilde{\tau} \geq u, A= a\}  \vert \bar{X}_{a}(u) \right]} &= \frac{\e_{0}\left[ \mathds{1}\{\tilde{\tau} \geq u, A= a\} H  \vert \bar{X}(u) \right]}{\e_{0}\left[ \mathds{1}\{\tilde{\tau} \geq u, A= a\}  \vert \bar{X}(u) \right]} \\
    &= \e_{0}\left[ H \vert \bar{X}(u), \tilde{\tau} \geq u, A= a \right].
\end{align*}
We recognize, from \eqref{etadef}, this conditional expectation as $\eta$ and we conclude that the integrand in the projection formula is 
\begin{align*}
   - \frac{\mathds{1}\{\tau \geq u, A=a\}}{\pi_{0}(a, W)G_{0}(u \vert \bar{X}, a)} \e_{0}\left[ H \vert \bar{X}(u), \tilde{\tau} \geq u, A= a \right].
\end{align*}
To see that $\eta$ as defined in \eqref{etadef} can be expressed as $\eta$ from the main text in \eqref{eq:Full_IF}, we will rewrite the conditional expectation in the expression \eqref{mainEtaId}. To that end, observe that by \eqref{WeightEq} we have 
\begin{align*}
     \e_{0}\left[\Delta_{\tau_{a},C_{a}}^{t_{0}} \frac{G_{0}(u \vert \bar{X}_{a},a)}{G_{0}(u \vee (\tau_{a} \wedge t_{0}) \vert \bar{X}_{a},a)}  \big\vert \bar{X}_{a}, C_{a} > u \right] = 1
\end{align*}
By iterated expectations we have 
\begin{align*}
    &\e_{0}\left[H_{a} \times 1 \big\vert \bar{X}_{a}(u), C_{a} > u \right] = \\
     &\e_{0}\left[ \e_{0}\left[ H_{a} \Delta_{\tau_{a},C_{a}} \frac{G_{0}(u \vert \bar{X}_{a},a)}{G_{0}(u \vee (\tau_{a} \wedge t_{0}) \vert \bar{X}_{a},a)}  \big\vert \bar{X}_{a}, C_{a} > u \right] \big\vert \bar{X}_{a}(u), C_{a} > u \right] = \\
     &\e_{0}\left[  H_{a} \Delta_{\tau_{a},C_{a}} \frac{G_{0}(u \vert \bar{X}_{a},a)}{G_{0}(u \vee (\tau_{a} \wedge t_{0}) \vert \bar{X}_{a},a)} \big\vert \bar{X}_{a}(u), C_{a} > u \right].
\end{align*}
Now note that in \eqref{WeightEq} we are multiplying by $\mathds{1}\{\tau_{a} \wedge t_{0} \geq u\}$ which is $\sigma(\bar{X}_{a}(u))$-measurable. On that event $u \vee (\tau_{a} \wedge t_{0}) = \tau_{a} \wedge t_{0}$. Therefore, on $\mathds{1}\{\tau_{a} \wedge t_{0} \geq u\}$ we have 
\begin{align*}
    \e_{0}\left[H_{a} \big\vert \bar{X}_{a}(u), C_{a} > u \right] = \e_{0}\left[  H_{a} \Delta_{T_{a},C_{a}} \frac{G_{0}(u \vert \bar{X}_{a})}{G_{0}(\tau_{a} \wedge t_{0} \vert \bar{X}_{a})} \big\vert \bar{X}_{a}(u), C_{a} > u \right].
\end{align*}

\newpage

\begin{lemma}
\label{lemma:iterexp_set}
    For any integrable random variable $B$ we have 
    \begin{align*}
        \e_{0}\left[ \e_{0}\left[B \Big\vert \bar{X}_{a}, C_{a} > u \right] \Big\vert \bar{X}_{a}(u), C_{a} > u \right] = \e_{0}\left[ B \Big\vert \bar{X}_{a}(u), C_{a} > u \right]
    \end{align*}
\end{lemma}
proof:
 \begin{align*}
        &\e_{0}\left[ \e_{0}\left[B \Big\vert \bar{X}_{a}, C_{a} > u \right]  \Big\vert \bar{X}_{a}(u), C_{a} > u \right] = \\
        &\frac{\e_{0}\left[ \mathds{1}\{C_{a} > u\}  \frac{\e_{0}\left[ \mathds{1}\{C_{a} > u\} B \Big\vert \bar{X}_{a} \right]}{\e_{0}\left[ \mathds{1}\{C_{a} > u\} \Big\vert \bar{X}_{a} \right]}  \Big\vert \bar{X}_{a}(u) \right]}{\e_{0}\left[ \mathds{1}\{C_{a} > u\} \Big\vert \bar{X}_{a}(u) \right]} = \\
        & \frac{\e_{0}\left[ \e_{0}\left[ \mathds{1}\{C_{a} > u\}  \frac{\e_{0}\left[ \mathds{1}\{C_{a} > u\} B \Big\vert \bar{X}_{a} \right]}{\e_{0}\left[ \mathds{1}\{C_{a} > u\} \Big\vert \bar{X}_{a} \right]} \Big\vert \bar{X}_{a} \right]  \Big\vert \bar{X}_{a}(u) \right]}{\e_{0}\left[ \mathds{1}\{C_{a} > u\} \Big\vert \bar{X}_{a}(u) \right]} = \\
        & \frac{\e_{0}\left[ \frac{\e_{0}\left[ \mathds{1}\{C_{a} > u\} B \Big\vert \bar{X}_{a} \right]}{\e_{0}\left[ \mathds{1}\{C_{a} > u\} \Big\vert \bar{X}_{a} \right]} \e_{0}\left[ \mathds{1}\{C_{a} > u\}   \Big\vert \bar{X}_{a} \right]  \Big\vert \bar{X}_{a}(u) \right]}{\e_{0}\left[ \mathds{1}\{C_{a} > u\} \Big\vert \bar{X}_{a}(u) \right]} = \\
        & \frac{ \e_{0}\left[ \mathds{1}\{C_{a} > u\} B \Big\vert \bar{X}_{a}(u) \right]  }{\e_{0}\left[ \mathds{1}\{C_{a} > u\} \Big\vert \bar{X}_{a}(u) \right]} = \\
        &\e_{0}\left[ B \Big\vert \bar{X}_{a}(u), C_{a} > u \right].
    \end{align*}

\subsection{ELOS and chain rule for IFs} 
\label{subsec:ELOS}

Another typical parameter of interest when working with multistate observations is, the \textit{expected length of stay} (ELOS), in various states, during some time interval. In our case, one could be interested in, say, differences in counterfactual ELOS between different baseline interventions. By definition, (counterfactual) ELOS from time zero to time $t$ is: 
\begin{align*}
    ELOS_{a}(t) &:= \e\left[ \int_{0}^{t} \mathds{1}\{Z_{a}(\tau_{a} \wedge u) = j\} \diff u \right] = \int_{0}^{t} \e\left[ \mathds{1}\{Z_{a}(\tau_{a} \wedge u) = j\} \right] \diff u.
\end{align*} 
We see that (counterfactual) ELOS, is a functional of the (counterfactual) state occupation probabilities. If this functional is differentiable (by a certain definition of differentiability), we can use the chain rule for influence functions \citep[Theorem 25.47]{vaart_1998} to derive the non-parametric efficient influence function for (counterfactual) ELOS as well. The necessary definition of smoothness for a chain rule to work, is \textit{compact differentiability}, or \textit{Hadamard differentiability} \citep[Chapter 20]{vaart_1998}. This generally produces a directional derivative, according to a chosen set of directions. The Hadamard derivative of the functional
\begin{align*}
    \rho(a) := \int a(u) \diff u
\end{align*}
in direction $\mathbb{H}$ for some suitable function space $\mathbb{H}$ of integrable functions (say, measurable and bounded) is
\begin{align*}
    \rho_{a}'(h) &=  \int h(u) \diff u.
\end{align*}
By the chain-rule for Hadamard differentiability (see \citet{vaart_1998} Theorem 20.9)), we have that the efficient influence function for (counterfactual) ELOS is given by
\begin{align*}
    IF_{\rho \circ \psi_{\obs} }(\obs,\theta)(t) &= \int_{0}^{t} IF_{\psi_{\obs}}(\obs,\theta)(s) \diff s,
\end{align*}
for $IF_{\psi_{\obs}}$ viewed as a function of $s$, replacing $t_{0}$. Using \eqref{eq:Full_IF}, we obtain that
\begin{align*}
    IF_{\rho \circ \psi_{\obs} }(\obs,\theta)(t) &= \frac{\mathds{1}\{A=a\}}{\pi(a,W)} \int_{0}^{t} \frac{H(s) \Delta^{s}}{G(\tilde{\tau} \wedge s  \vert \bar{X}(C),A)} \diff s \ + \\
    &\frac{\mathds{1}\{A=a\}}{\pi(a,W)} \int_{0}^{t} \int_{0}^{\tilde{\tau} \wedge s} \frac{\eta(s,u,\bar{X}(u),a)}{G(u  \vert \bar{X}(C),A)} \diff M^{G}(u) \diff s  \ - \\
    &\left(\frac{\mathds{1}\{A=a\}}{\pi(a,W)} - 1\right) \int_{0}^{t} Q(s,a,W) \diff s - ELOS_{a}(t).
\end{align*}

Thus, we see that estimation of ELOS follows from the estimation of occupation probabilities. For the remainder of the paper, we focus only on estimators for occupation probabilities.
\newpage

\appendixthree

\section{(Robustness and modified estimators)} 
\label{app:C:Robust_and_mod_est}

\subsection{Parameter robustness}

Recall that by robustness of a target estimator, say $\hat{\psi}_{n}^{\mu}$, we here mean conditions under which the parameter function $\mu$ as a function of nuisance-parameters $\gamma$, is unbiased. The robustness claims for the IPW estimator is clear. We prove the robustness claims of Section \eqref{subsec:prop} for $\hat{\psi}_{n}^{\nu}$ and $\hat{\psi}_{n}^{\mu}$. Let $\gamma_{1} = (\pi_{1}, G_{1}, \eta_{1}, Q_{1})$ which we interpret as the limit, in probability, of nuisance parameter estimators and let $\gamma_{0} = (\pi_{0}, G_{0}, \eta_{0}, Q_{0})$. Recall that we defined in \ref{eq:nu} and \ref{eq:mu}:
\begin{align*}
\nu(\obs, \gamma_{1}) &= \frac{\mathds{1}\{A = a\} \Delta^{t_{0}} H}{\pi_{1}(a, W)G_{1}(\tilde{\tau} \wedge t_{0} \vert \bar{X}, a)}  - \left(\frac{\mathds{1}\{A=a\} - \pi_{1}(a, W)}{\pi_{1}(a, W)}\right)Q_{1}(a,W). \\
\mu(\obs, \gamma_{1}) &= \nu(\obs, \gamma_{1}) + \frac{\mathds{1}\{A = a\} }{\pi_{1}(a, W)} \int_{0}^{\tilde{\tau} \wedge t_{0}} \frac{\eta_{1}(u, \bar{X}(u), a)}{G_{1}(u \vert \bar{X}, a)} \diff M^{G_{1}} (u).
\end{align*}

To prove the claims we first show that $\nu$ is unbiased if $\pi_{1} = \pi_{0}$ or $Q_{1} = Q_{0}$. Then we show that the martingale term has expectation zero if $G_{1} = G_{0}$ or $\eta_{1} = \eta_{0}$. For the first term of $\nu(\obs, \gamma_{1})$, we have
\begin{align*}
    \e_{0}\left[\frac{\mathds{1}\{A = a\} \Delta^{t_{0}} H}{\pi_{1}(a, W)G_{1}(\tilde{\tau} \wedge t_{0} \vert \bar{X}, a)} \right] &= \e_{0}\left[\e_{0}\left[ \e_{0}\left[\frac{\mathds{1}\{A = a\} \Delta_{a}^{t_{0}}  H_{a} }{\pi_{1}(a, W)G_{1}(\tilde{\tau} \wedge t_{0} \vert \bar{X}, a)} \vert \bar{X}_{a} \right] \vert W \right]  \right]\\
    &= \e_{0}\left[\frac{\pi_{0}(a, W)  }{\pi_{1}(a, W)} \e_{0}\left[ H_{a} \vert W \right]\right] =  \e_{0}\left[\frac{\pi_{0}(a, W) }{\pi_{1}(a, W)} Q_{0}(a, W) \right].\\
\end{align*}
The second we get, using iterated expectations conditioning on $W$,
\begin{align*}
     \e_0 \Bigg[ \Bigg(\frac{\mathds{1}\{A=a\} - \pi_{1}(a, W)}{\pi_{1}(a, W)}\Bigg)Q_{1}(a,W) \Bigg]  &= \e_0 \Bigg[ \Bigg(\frac{ \overbrace{\e_0 \left[ \mathds{1}\{A=a\} \mid W \right]}^{\pi_0(a,W)}  - \pi_{1}(a, W)}{\pi_{1}(a, W)}\Bigg)Q_{1}(a,W) \Bigg] 
\end{align*}
Therefore, 
\begin{align*}
    \e_{0}[\nu(\obs,\gamma_{1})] &=\e_{0}\left[  \frac{\pi_{0}(a, W)}{\pi_{1}(a, W)} (Q_{0}(a, W) - Q_{1}(a, W)) + Q_{1}(a, W)\right]
\end{align*}
which equals $\psi_{0}$ if $\pi_{1} = \pi_{0}$ or $Q_{1} = Q_{0}$.

\noindent Now we consider the martingale term. We first prove a useful identity for $G$-weighted martingale. Here we argue as in \cite{Scheike2020} appendix B (see also fundamental identities of \cite{Robins1992}). First, we note that for any coarsening kernel $G$ (again suppressing index notation), with corresponding cumulative hazard function $\Lambda^{G}$, we have
\begin{align}
    \int_{0}^{\tilde{\tau} \wedge t_{0}} \frac{1}{G(u \vert \bar{X}, A)}  \diff M^{G}(u) &= 1 - \frac{\mathds{1}\{C > \tilde{\tau} \wedge t_{0}\}}{G(\tilde{\tau} \wedge t_{0} \vert \bar{X}, A)} \label{imp_eq}
\end{align}
To see this first of all note that 
\begin{align*}
    \int_{0}^{\tilde{\tau} \wedge t_{0}}\frac{1}{G(u \vert \bar{X}, A)} \diff \Lambda^{G}(u \vert \bar{X}, A)   &= \int_{0}^{\tilde{\tau} \wedge t_{0}} \exp\left( \Lambda^{G}(u \vert \bar{X}, A) \right) \diff \Lambda^{G}(u \vert \bar{X}, A) \\
    &= \exp\left( \Lambda^{G}(\tilde{\tau} \wedge t_{0} \vert \bar{X}, A) \right) - \exp\left( \Lambda^{G}(0 \vert \bar{X}, A) \right)  \\
    &= \frac{1}{G(\tilde{\tau} \wedge t_{0} \vert \bar{X}, A)} - 1
\end{align*}
where the second equality follows by the chain rule. From the counting process part of the martingale we get 
\begin{align*}
    \int_{0}^{\tilde{\tau} \wedge t_{0}} \frac{1}{G(u \vert \bar{X}, A)}  \diff N^{C} &=  \frac{\mathds{1}\{C \leq \tilde{\tau} \wedge t_{0}\}}{G(\tilde{\tau} \wedge t_{0} \vert \bar{X}, A)} =  \frac{1 - \mathds{1}\{C > \tilde{\tau} \wedge t_{0}\}}{G(\tilde{\tau} \wedge t_{0} \vert \bar{X}, A)}
\end{align*}
and \eqref{imp_eq} follows. Suppose that $\eta_{1} = \eta_{0}$. We will use the representation \eqref{etadef} i.e. $\eta_{0}(u, \bar{X}(u), a) = \e_{0}[H \vert \bar{X}(u), \tilde{\tau} > u, A= a]$, justified in Appendix \eqref{app:B:inf_fn}. By the identity \eqref{imp_eq} we have that
\begin{align*}
    \frac{H \cdot \overbrace{\mathds{1}\{C > \tilde{\tau} \wedge t_{0}\}}^{\Delta^{t_{0}} :=}}{G_{1}(\tilde{\tau} \wedge t_{0} \vert \bar{X}, A)} & = H \cdot \left( 1 - \left( 1 -\frac{ \mathds{1}\{C > \tilde{\tau} \wedge t_{0}\})}{G_{1}(\tilde{\tau} \wedge t_{0} \vert \bar{X}, A)} \right) \right) \\ 
    & \stackrel{\eqref{imp_eq}}{=} H - \int_{0}^{\tilde{\tau} \wedge t_{0}} \frac{H}{G_{1}(u \vert \bar{X}, A)}  \diff M^{G_{1}}(u)
\end{align*}
and it follows that:
\begin{align*}
    \frac{H\Delta^{t_{0}}}{G_{1}(\tilde{\tau} \wedge t_{0} \vert \bar{X}, A)} + \int_{0}^{\tilde{\tau}\wedge t_{0}} \frac{\eta_{1}(u, \bar{X}(u), A)}{G_{1}(u \vert \bar{X}, A)} \diff M^{G_{1}} &= H + \int_{0}^{\tilde{\tau}\wedge t_{0}} \frac{(\eta_{1}(u, \bar{X}(u), A) - H)}{G_{1}(u \vert \bar{X}, A)} \diff M^{G_{1}}
\end{align*}
If we let $\epsilon(u) := \frac{(\eta_{1}(u \vert \bar{X}(u), A) - H)}{G_{1}(u \vert \bar{X}, A)}$. We see that
\begin{align*}
    \int_{0}^{\tilde{\tau}\wedge t_{0}} \epsilon(u)  \diff M^{G_{1}} &= \int_{0}^{\tilde{\tau}\wedge t_{0}} \epsilon(u)  \diff M^{G_{0}} + \int_{0}^{\tilde{\tau}\wedge t_{0}} \epsilon(u) \mathds{1}\{\tilde{\tau} \geq u\}  \diff (\Lambda^{G_{0}} - \Lambda^{G_{1}})
\end{align*}
The first term has conditional mean zero under $\prp_{0}$. We will show that the second term has conditional mean zero given $A,W$. We have 
\begin{align*}
    \e_{0}[\epsilon(u) \vert \bar{X}(u), \tilde{\tau} > u, A, W] &= \frac{\e_{0}[\eta_{1}(u, \bar{X}(u), A) - H \vert \bar{X}(u), \tilde{\tau} > u, A, W]}{G_{1}(u \vert \bar{X}, A)} \\
    &= \frac{\eta_{1}(u, \bar{X}(u), A) - \eta_{0}(u, \bar{X}(u), A)}{G_{1}(u \vert \bar{X}, A)} = 0.
\end{align*}
Here the first equality, taking the kernel outside of the conditional expectation, is due to CAR. The second is by assumption that $\eta_{1} = \eta_{0}$. Now conditional Fubini gives us 
\begin{align*}
    \e_{0}\left[\int_{0}^{ t_{0}} \epsilon(u) \mathds{1}\{\tilde{\tau} \geq u\}  \lambda^{G}(u)  \diff u \Big\vert A, W\right] &= \int_{0}^{  t_{0}} \e_{0}\left[ \epsilon(u) \mathds{1}\{\tilde{\tau} \geq u\}  \lambda^{G}(u) \Big\vert A, W\right] \diff u.
\end{align*}
Iterated expectations gives us 
\begin{align*}
    &\e_{0}\left[ \epsilon(u) \mathds{1}\{\tilde{\tau} \geq u\}  \lambda^{G}(u) \Big\vert A, W\right] \\
    & = \e_{0}\left[ \e_{0}\left[ \epsilon(u) \mathds{1}\{\tilde{\tau} \geq u\}  \lambda^{G}(u) \Big\vert \bar{X}(u), \tilde{\tau} > u, A, W \right]  \Big\vert A, W\right] \\
    &= \e_{0}\Big [ \underbrace{\e_{0}\left[ \epsilon(u) \Big\vert \bar{X}(u), \tilde{\tau} > u, A, W \right]}_{= \ 0 \text{, shown above }}  \mathds{1}\{\tilde{\tau} \geq u\}  \lambda^{G}(u) \Big\vert A, W\Big] = 0
\end{align*}
This holds for any $\lambda^{G}$, that is for any $G$ with hazard $\lambda^{G}$ therefore, the martingale term has expectation zero. If $G_{1} = G_{0}$ the censoring martingale is a mean zero martingale. Since the tangent space $\mathbb{T}_{G}$ restricts to functions $\eta$ such that the martingale integral is zero, we may without loss of generality, restrict the domain of $\eta$ parameters to such $\eta$. Doing so implies the martingale term has mean zero for $G_{1} = G_{0}$.

\newpage

\appendixfour

\section{(Convergence rates)} 
\label{app:D:rates}
\subsection{The $\mu$-estimator}

We first show the representation \eqref{remainder} of the remainder term between deterministic functions $\mu(\cdot) = \mu(\cdot, \gamma)$ and $\mu_{0}(\cdot) = \mu(\cdot, \gamma_{0})$. To that end we use a different representation of $\mu$ (and similarly for $\mu_{0}$) using equality \eqref{app:C:Robust_and_mod_est}. We have that
\begin{align}
        \mu(\obs, \gamma) &= \frac{\mathds{1}\{A=a\} \Delta}{\pi G}H - \left( \frac{\mathds{1}\{A=a\}}{\pi} - 1 \right)Q + \frac{\mathds{1}\{A=a\} }{\pi } \int \frac{\eta}{G} \diff M^{G} \nonumber \\
        &= \frac{\mathds{1}\{A=a\} }{\pi } \left( \frac{\Delta}{G}H - Q \right) + Q + \frac{\mathds{1}\{A=a\} }{\pi } \int \frac{\eta}{G} \diff M^{G} \nonumber \\
        &= \frac{\mathds{1}\{A=a\} }{\pi } \left( H - \int \frac{H}{G} \diff M^{G} - Q \right) + Q  + \frac{\mathds{1}\{A=a\} }{\pi } \int \frac{\eta}{G} \diff M^{G} \nonumber \\
        &= \frac{\mathds{1}\{A=a\} }{\pi } (H - Q) + Q + \frac{\mathds{1}\{A=a\} }{\pi } \int \frac{(\eta-H)}{G} \diff M^{G} \label{mu_rep}
    \end{align}

Using \eqref{mu_rep} and iterated expectations analogous to the argumentation in appendix \ref{app:C:Robust_and_mod_est} we get the following remainder term
\begin{align}
    R(\mu, \mu_{0}) &= E_{\theta_{0}}\left[ \mu(O, \gamma) - \mu(O, \gamma_{0}) \right] \nonumber \\
    &= E_{\theta_{0}}\left[ \mu(O, \gamma) - Q_{0} \right] \pm E_{\theta_{0}}\left[ \frac{\mathds{1}\{A=a\}}{\pi} \int\frac{(\eta - H)}{G} \mathrm{d}\Lambda^{G_{0}} \right] \nonumber \\
    &=  E_{\theta_{0}}\left[ \frac{\mathds{1}\{A=a\} }{\pi } (H - Q) + Q - Q_{0} \right]  + E_{\theta_{0}}\left[ \frac{\mathds{1}\{A=a\}}{\pi} \int\frac{(\eta - H)}{G} \mathrm{d}(\Lambda^{G_{0}} -\Lambda^{G}) \right] \nonumber \\
    &= E_{\theta_{0}}\left[ \left(\frac{\pi_{0}}{\pi} - 1 \right) \left( Q_{0}(G_{0}) - Q(G) \right) \right] \ + \label{determ_remainder} \\
    &E_{\theta_{0}}\left[ \left(\frac{\pi_{0}}{\pi} - 1 \right) \int\frac{(\eta(G) - \eta_{0}(G_{0})}{G} \mathrm{d}(\Lambda^{G_{0}} -\Lambda^{G}) \right] + \nonumber  \\
    &E_{\theta_{0}}\left[\int\frac{(\eta(G) - \eta_{0}(G_{0})}{G} \mathrm{d}(\Lambda^{G_{0}} -\Lambda^{G}) \right] \nonumber .
\end{align}
To establish a similar representation of the remainder term in the case of a random function $\hat{\mu}(\cdot) = \mu(\cdot, \hat{\gamma})$ for $\hat{\gamma}$ based on observed data $\bar{O}_{n} = (O_{1}, \ldots, O_{n})$ we use a similar representation of $\mu$ as the one established in \eqref{mu_rep}. For that we need a similar representation as the weighted martingale in \eqref{imp_eq}. To that end, consider a generic individual with observed data $O$ and let $\hat{\Lambda}$ be the cumulative censoring hazard evaluated at this data point based on the Breslow estimator. The estimated censoring survival is then
\begin{align*}
    \hat{G}_{n}(t) = \prod_{j: C_{j} \wedge \tilde{\tau}}\left(1 - \Delta\hat{\Lambda}_{n}(C_{j}) \right).
\end{align*}
We have that
\begin{align*}
    \frac{1}{\hat{G}_{n}(t)} = \prod_{j: C_{j} \leq t\wedge \tilde{\tau}} \frac{1}{\left(1 - \Delta\hat{\Lambda}_{n}(C_{j}) \right)} = \prod_{j: C_{j} \leq t\wedge \tilde{\tau}} \left(1 + \Delta\hat{U}_{n}(C_{j}) \right)
\end{align*}
for a piecewise constant right-continuous function function $\hat{U}_{n}$ with jumps
\begin{align*}
    \Delta\hat{U}_{n} = \frac{\Delta\hat{\Lambda}_{n}}{1-\Delta\hat{\Lambda}_{n}}.
\end{align*}
By \cite{gill90} Theorem 5, the product integral based on $\hat{U}_{n}(t)$ satisfies the forward equation 
\begin{align*}
    \prod_{(0,t]}\left(1 + \mathrm{d}\hat{U}_{n}(s) \right) - 1&= \int \prod_{(0,u)}\left(1 + \mathrm{d}\hat{U}_{n}(s) \right) \mathrm{d}\hat{U}_{n}(u)
\end{align*}
That is
\begin{align*}
    \frac{1}{\hat{G}_{n}(t)} - 1 &= \int \frac{1}{\hat{G}_{n}(u-)(1-\Delta\hat{\Lambda}_{n}(u))} \mathrm{d}\hat{\Lambda}_{n,i}(u) = \int \frac{1}{\hat{G}_{n}(u)} \mathrm{d}\hat{\Lambda}_{n}(u)
\end{align*}
Therefore, we get an analogous representation as \eqref{mu_rep} for $\mu(o, \hat{\gamma})$. Let $\hat{\mu} := \mu(o, \hat{\gamma})$ recall that $R(\hat{\mu}, \mu_{0}) := P_{0}\{\hat{\mu} - \mu_{0}\}$. Let $O \sim P_{0}$ be independent of $\bar{O}_{n}$, then we have the convenient representation
\begin{align*}
    P_{0}\{\hat{\mu} - \mu_{0}\} &= E[ \mu(O, \hat{\gamma}) - \mu(O, \gamma_{0}) \ \vert \ \bar{O}_{n} ]
\end{align*}
Using the same line of arguments as in leading to \eqref{determ_remainder} along with iterated expectations arguments of appendix \ref{app:C:Robust_and_mod_est}, condoning additionally on parts of the history of $O$ we arrive at  
\begin{align*}
    R(\hat{\mu}, \mu_{0}) &= P_{0}\left[ \left(\frac{\pi_{0}}{\hat{\pi}} - 1 \right) \left( Q_{0}(G_{0}) - \hat{Q}(\hat{G}) \right) \right] + \\
    &P_{0}\left[ \left(\frac{\pi_{0}}{\hat{\pi}} - 1 \right) \int\frac{(\hat{\eta}(\hat{G}) - \eta_{0}(G_{0})}{\hat{G}} \mathrm{d}(\Lambda^{G_{0}} -\hat{\Lambda}^{G}) \right] + \\
    &P_{0}\left[\int\frac{(\hat{\eta}(\hat{G}) - \eta_{0}(G_{0})}{\hat{G}} \mathrm{d}(\Lambda^{G_{0}} -\hat{\Lambda}^{G}) \right]
\end{align*}
A simple bound on the remainder term is obtained by applications of Cauchy-Schwarz along with positivity constraints on $\hat{\pi}$ and $\hat{G}$ leading to 
\begin{align*}
    \vert R(\hat{\mu}, \mu_{0}) \vert \lesssim  \Vert \pi_{0} - \hat{\pi} \Vert \Vert Q_{0}(G_{0}) - \hat{Q}(\hat{G}) \Vert + \left\{ \Vert \pi_{0} - \hat{\pi} \Vert + 1 \right\} \left\Vert \int (\hat{\eta}(\hat{G}) - \eta_{0}(G_{0}) \mathrm{d}(\Lambda^{G_{0}} -\hat{\Lambda}^{G})  \right\Vert
\end{align*}
Due to assumption \eqref{MG_assumption} and $\Vert \pi_{0} - \hat{\pi} \Vert =O_{p}(1)$ it suffices to show that 
\begin{align*}
    \Vert \pi_{0} - \hat{\pi} \Vert \Vert Q_{0}(G_{0}) - \hat{Q}(\hat{G}) \Vert = o_{p}(1/ \sqrt{n})
\end{align*}
First, by the triangle inequality we further bound the product term by 
\begin{align*}
    \Vert \pi_{0} - \hat{\pi}  \Vert  \left( \Vert Q_{0}(G_{0}) - \hat{Q}(G_{0}) \Vert + \Vert \hat{Q}(G_{0}) - \hat{Q}(\hat{G}) \Vert \right)
\end{align*}
Since $\Vert Q_{0}(G_{0}) - \hat{Q}(G_{0}) \Vert = O_{P}(1/r_{n}^{\hat{Q}})$, $\Vert \pi_{0} - \hat{\pi}  \Vert = O_{P}(1/r_{n}^{\hat{\pi}})$ and $\Vert \hat{Q}(G_{0}) - \hat{Q}(\hat{G}) \Vert = O_{P}(1/r_{n}^{\hat{G}})$, the rate constraint \eqref{product_rate_condition} of the main text conclude the proof.

\newpage

\appendixfive

\section{Extended Simulation Results} 
\label{app:E:simresults}

\subsection{Additional Details on Scenario Design}
\label{app:E:subsec:scendesign}

\begin{figure}[H]
    \centering
    \includegraphics[width=\linewidth]{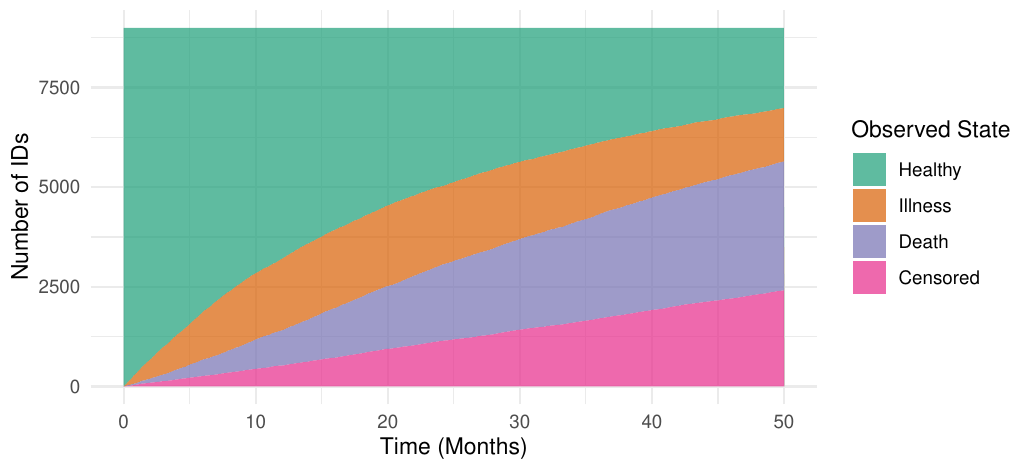}
    \caption{The observed state occupation process, demonstrated on simulation dataset ID $\#1$, with $9000$ observations.}
    \label{fig:observed_stackplot}
\end{figure}

\begin{itemize}
    \item The joint distribution of \textit{Sex} and \textit{Age} is simulated from the 2025 population distribution across the country of Norway, sourced from the Norwegian statistics bureau SSB\footnote{\url{https://www.ssb.no/en/statbank/table/07459/}}, and filtered for ages between 16 and 95. \textit{BMI} is simulated as a $N(26.5, 5^2)$ normal distribution truncated between 10 and 40 and floored to integer values. Truncation of \textit{Age} and \textit{BMI} allows us to bound propensity scores and censoring probability curves when setting coefficients to ensure positivity holds in the full population.
    \item  We use sigmoid transformations of the \textit{Age} and \textit{BMI} variables to capture non-linear impact on risks. Risk contribution from \textit{BMI} is modelled with piecewise sigmoids mirrored around $21.5$, with means $-17$ and $26$, a slope of $1.1$ and scaled by $0.75$. For \textit{Age}, we use a sigmoid centered at $60$ with slope parameter $0.15$, both from the \texttt{pracma} package. See Figure \ref{fig:BMI_Age_transf} below for a visual representation of the functions $f_1$ and $f_2$.
    \begin{figure}[H]
    \centering
    \includegraphics[width=\linewidth]{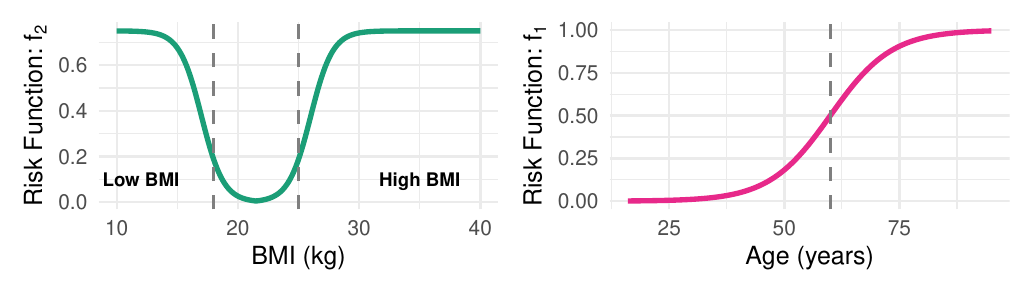}
    \caption{$f_1$ and $f_2$ risk transformations of \textit{Age} and \textit{BMI}.}
    \label{fig:BMI_Age_transf}
\end{figure}
    \item The state transition hazards $\alpha_{ij}$ in the CTMC are linear sums of \textit{A}, \textit{Sex}, $f_1(\textit{Age})$ and $f_2(\textit{BMI})$, with interaction terms between \textit{A} and the other variables. Overall, this means that there are $32$ coefficients chosen:
    \begin{align*}
          \begin{bmatrix}
        \alpha_{12} \\
        \alpha_{13} \\
        \alpha_{21} \\
        \alpha_{23} 
    \end{bmatrix} =
    \left[
\begin{array}{rrrrrrrr}
    -3.950 & -1.20 &  0.60 &  0.80 & 0.80 & -0.30 & -0.30 & -0.30 \\
    -5.550 & -1.20 &  0.60 &  0.80 & 0.80 & -0.30 & -0.30 & -0.30 \\
    -5.150 & -1.80 & -0.60 & -0.80 & - 0.80 &  0.30 &  0.30 &  0.30 \\
    -4.200 & -1.20 &  0.60 &  0.80 & 0.80 & -0.30 & -0.30 & -0.30 
\end{array} \right]
    \begin{bmatrix}
    \underline{X} \\
    \underline{X}  \\
    \underline{X}  \\
    \underline{X}  
    \end{bmatrix}, \text{ where} \\
    \underline{W} := ( \textit{Sex}, f_1(\textit{Age}), f_2(\textit{BMI})), \; \text{ and } \; \underline{X} := (1, A, \underline{W}, A\cdot \underline{W}) \in \mathbb{R}^8.
    \end{align*}
    The coefficients were set to have a dynamic where the hypothetical treatment is significantly beneficial, but the occupation probability of both the treated and non-treated populations are still bounded from 0\% and 100\% for $t>0$. We ensure that this also holds for individuals with all, but the most extreme covariate value combinations, as demonstrated in Figure \ref{fig:true_occprob}.
    
\item The coefficients for treatment assignment from equation \eqref{eq:pi_correct_form} are $\xi = (-2.10, 2.50, 2.50)$. These ensure a propensity score highly dependent on individual covariates, where the positivity assumption \eqref{eq:ObsPos} still holds, with values roughly bounded in the interval $[0.10, 0.90]$ as shown in Figure \ref{fig:true_propensity} below. We simulate only on the interval $[0,50]$.

\begin{figure}[H]
    \centering
    \includegraphics[width=\linewidth]{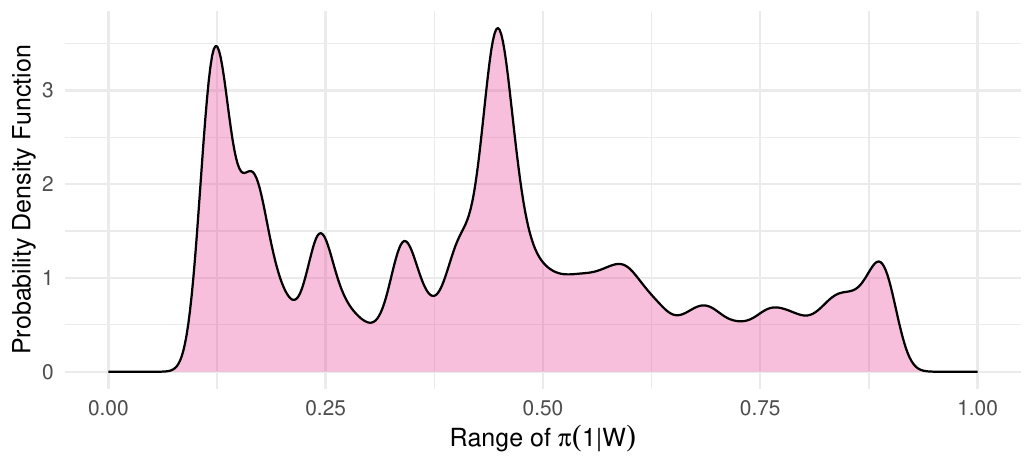}
    \caption{Propensity score distribution in our example scenario. We simulated $9000$ individuals from the joint covariate distribution and plotted the histogram of their propensity scores using the true coefficients.}
    \label{fig:true_propensity}
\end{figure}

\item The coefficients for the censoring hazard model from equation \eqref{eq:G_correct_form} are
\begin{align*}
    \beta = \left( e^{-6.50}, 0.80, 0.75, 0.75, 0.75, \frac{1.10}{25} \right).
\end{align*}
The coefficients chosen ensure a highly dependent censoring-scheme, where $P(C > t | W, \bar{Z}_A(t))$ is bounded from $0$ for even the most extreme covariate value combinations and transition histories.
\end{itemize}
Overall, the observed (biased) occupation probability in the simulated datasets is presented in Figure \ref{fig:observed_stackplot}.

\newpage

\subsection{Alternative test for robustness to  $G_0$ estimation}
\label{app:subsec:moreplots}

We define the biased estimator $\hat \lambda^G_{1b}$ of $\lambda^{G_0}$ by completely removing the \textit{Age} and \textit{BMI} variables:
\begin{align}
   \hat \lambda^G_{1b}(t) & := \hat \gamma_0 \exp (\hat  \gamma_1 A + \hat \gamma_2 \mathds{1}\{Z(t-) = 2\} + \hat \gamma_3 \tau_{12}(t-)). \label{eq:G_1b}
\end{align}

\begin{figure}[H]
    \centering
    \includegraphics[width=\linewidth]{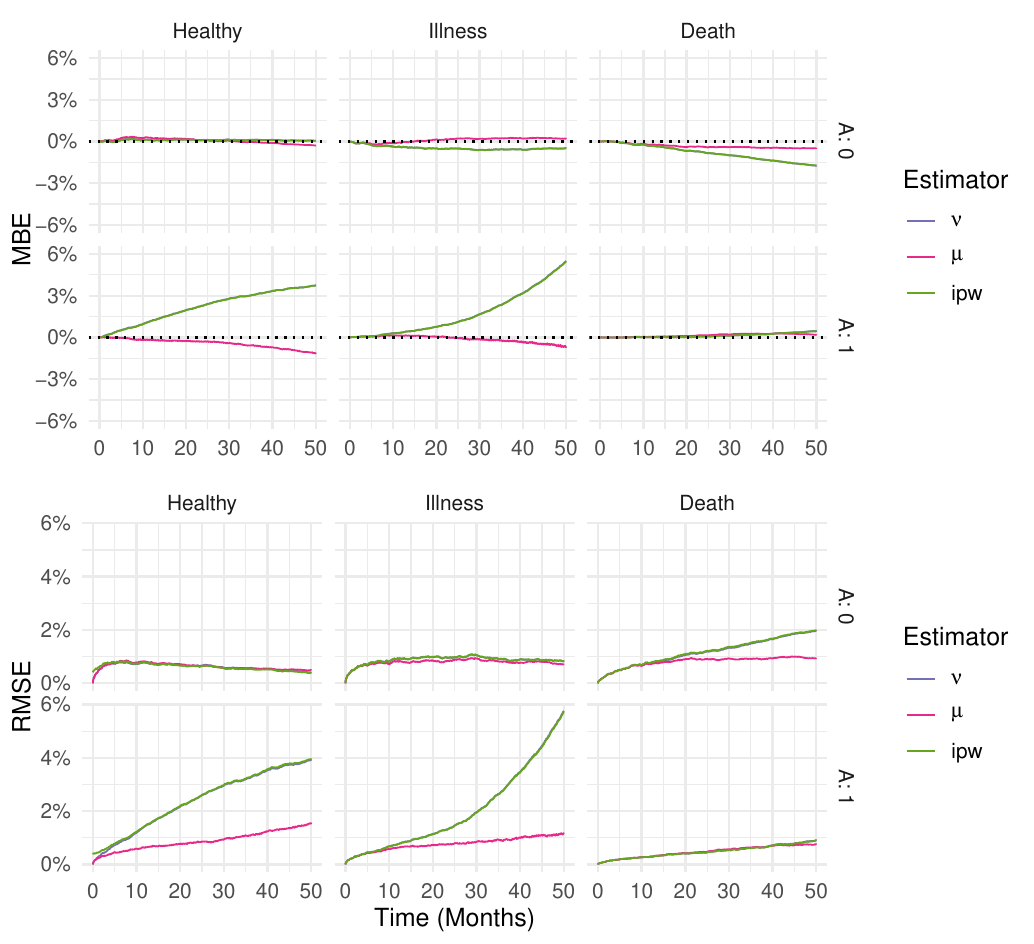}
    \caption{MBE and RMSE metrics of the estimators $\hat{\psi}_{n}^{0}$, $\hat{\psi}_{n}^{\nu}$, $\hat{\psi}_{n}^{\mu}$, $\hat{\psi}_{n}^{\nu, mod}$ and $\hat{\psi}_{n}^{\mu, mod}$, all using the biased nuisance estimator $\hat \lambda^G_{1b}$ \eqref{eq:G_1b}. Averages are over $100$ independent datasets with a population size of $9000$ in each.}
    \label{fig:Test3c_MBE_RMSE}
\end{figure}

\newpage

\subsection{Robustness to $Q_0$ and $\eta_0$ estimation}
\label{subsec:sens_to_Q_eta}

\begin{figure}[H]
    \centering
    \includegraphics[width=\linewidth]{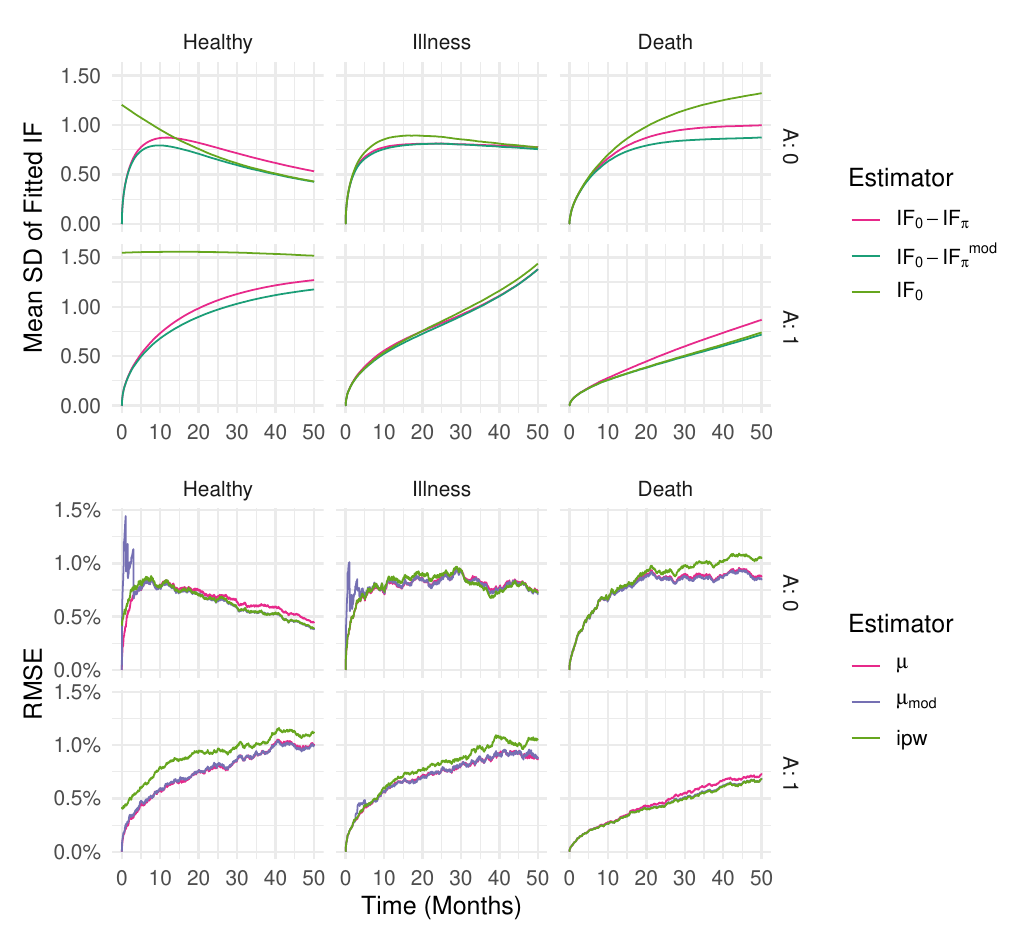}
    \caption{\textbf{Robustness to $Q$}: Mean SD and RMSE metrics of the estimators $\hat{\psi}_{n}^{0}$, $\hat{\psi}_{n}^{\mu}$ and $\hat{\psi}_{n}^{\mu, mod}$, all using the biased nuisance estimator $\hat Q_1(t)$. Averages are over $100$ independent datasets with a population size of $9000$ in each.}
    \label{fig:Test4a_MSD_RMSE}
\end{figure}

\newpage

\begin{figure}[H]
    \centering
    \includegraphics[width=\linewidth]{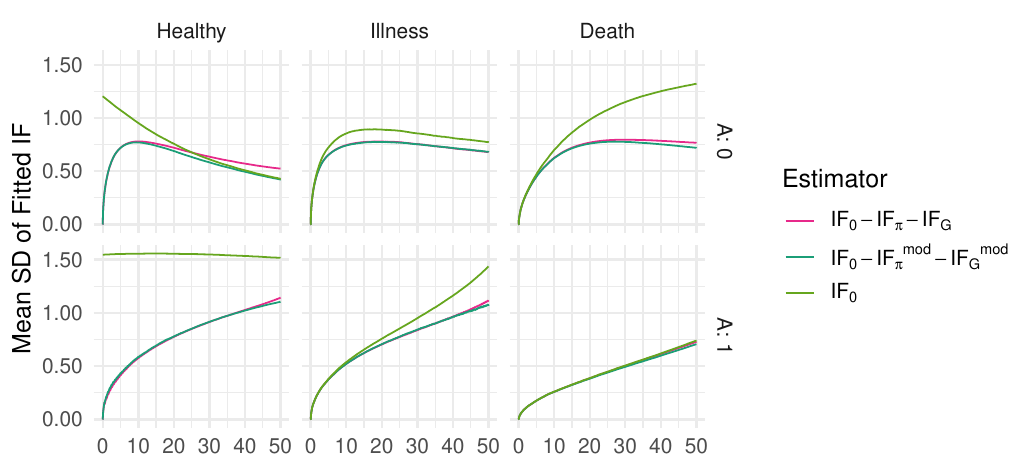}
    \caption{\textbf{Robustness to $\eta$}: Mean SD of the influence functions for estimators $\hat{\psi}_{n}^{0}$, $\hat{\psi}_{n}^{\mu}$ and $\hat{\psi}_{n}^{\mu, mod}$, all using the biased nuisance estimator $\hat \eta_1(t,u)$. Averages are over $100$ independent datasets with a population size of $9000$ in each.}
    \label{fig:Test4b_MSD}
\end{figure}

\newpage

\subsection{Diagnostics from modified estimator coefficients}
\label{app:mod_coef_diagnostics}

\begin{figure}[H]
    \centering
    \includegraphics[width=\linewidth]{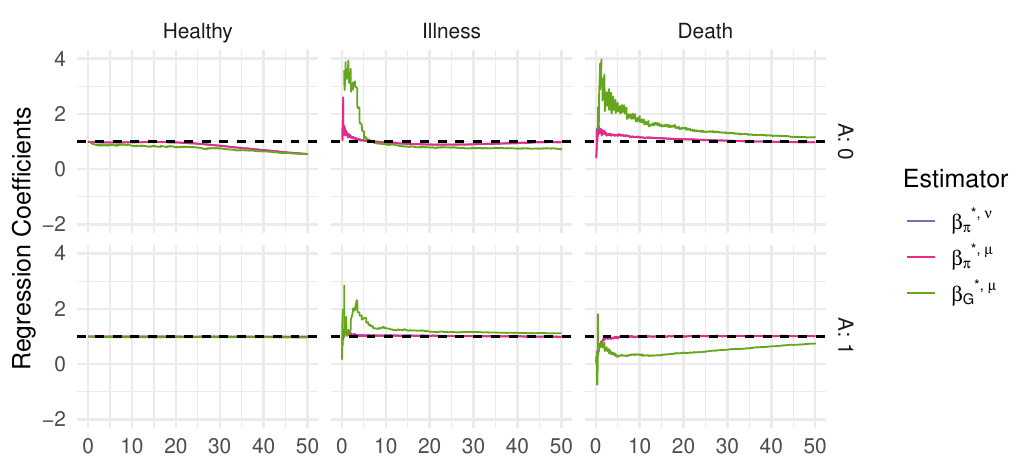}
    \caption{Modifying regression coefficients of $\hat{\psi}_{n}^{\nu, mod}$ and $\hat{\psi}_{n}^{\mu, mod}$ when using the consistent nuisance parameter estimators $\hat \pi_0$, $\hat \lambda^{G}_0$, $\hat Q_0$ and $\hat \eta_0$. Averages are over $100$ independent datasets with a population size of $9000$ in each.}
    \label{fig:Test1_ModCoefs}
\end{figure}

\begin{figure}[H]
    \centering
    \includegraphics[width=\linewidth]{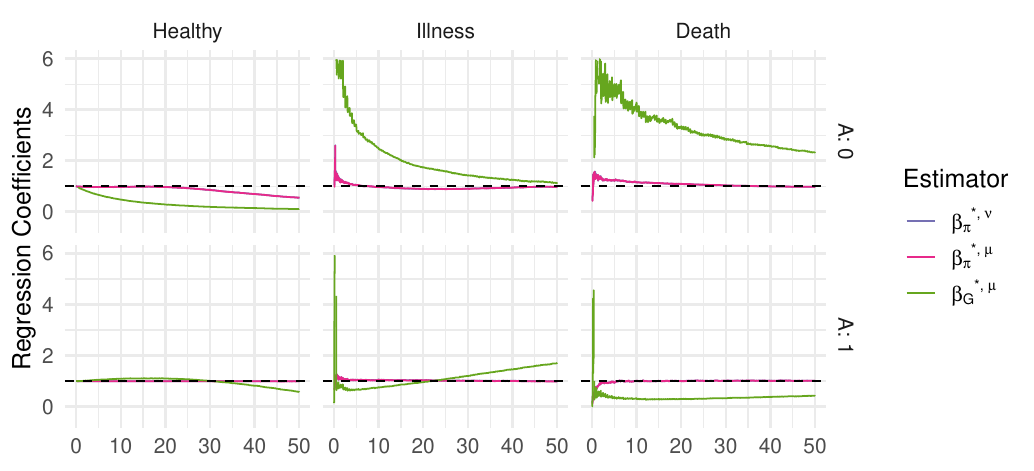}
    \caption{\textbf{Robustness to $\eta$}: Modifying regression coefficients of $\hat{\psi}_{n}^{\nu, mod}$ and $\hat{\psi}_{n}^{\mu, mod}$ when using the biased nuisance estimator $\hat \eta_1(t,u)$. Averages are over $100$ independent datasets with a population size of $9000$ in each.}
    \label{fig:Test4b_ModCoefs}
\end{figure}

\newpage

\begin{figure}[H]
    \centering
    \includegraphics[width=\linewidth]{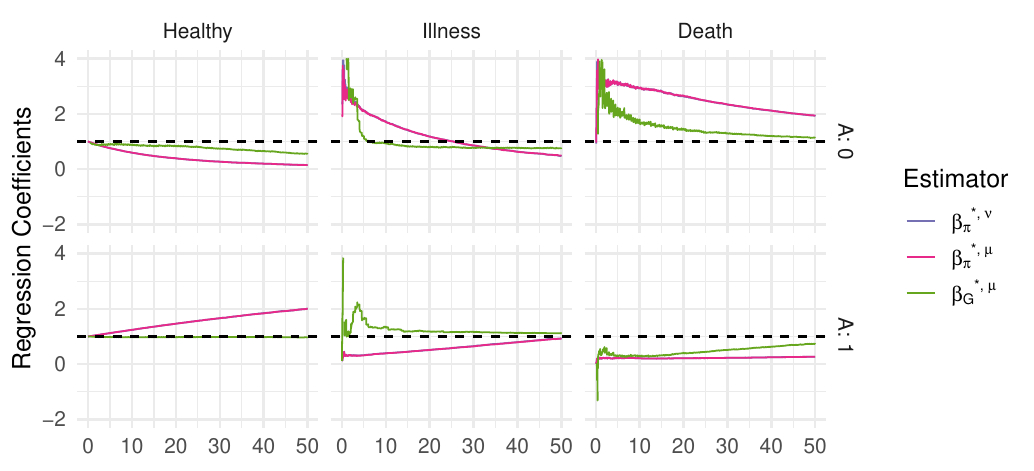}
    \caption{\textbf{Robustness to $Q$}: Modifying regression coefficients of $\hat{\psi}_{n}^{\nu, mod}$ and $\hat{\psi}_{n}^{\mu, mod}$ when using the biased nuisance estimator $\hat Q_1(t)$. Averages are over $100$ independent datasets with a population size of $9000$ in each.}
    \label{fig:Test4a_ModCoefs}
\end{figure}

\newpage

\begin{figure}[H]
    \centering
    \includegraphics[width=\linewidth]{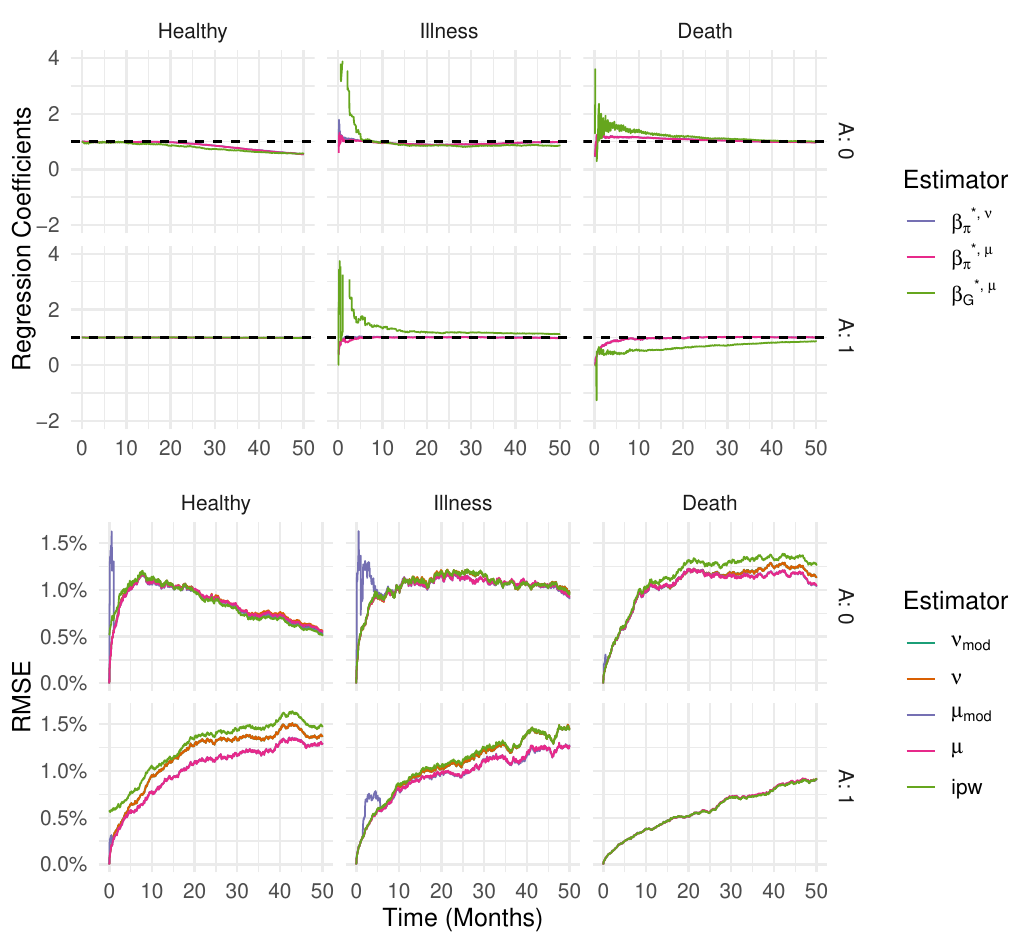}
    \caption{RMSE of the five estimators and the modifying regression coefficients of $\hat{\psi}_{n}^{\nu, mod}$ and $\hat{\psi}_{n}^{\mu, mod}$ when using the consistent nuisance parameter estimators $\hat \pi_0$, $\hat \lambda^{G}_0$, and $\hat Q_0$, but adding $A$ interaction terms to $\hat \eta_0$. Averages are over $100$ independent datasets with a population size of $5000$ in each.}
    \label{fig:Test5_RMSE_ModCoefs}
\end{figure}

\end{document}